\documentclass[showpacs,nofootinbib,amsmath,amssymb,superscriptaddress,unsortedaddress]{revtex4}
\usepackage[latin1]{inputenc}
\usepackage[T1]{fontenc}
\usepackage{graphicx}
\usepackage{epsfig}

\usepackage{hyperref}

\begin{document}

\newcommand{\mhalfo}{\frac{1}{2}}	
\newcommand{\mhalf}[1]{\frac{#1}{2}}
\newcommand{\ka}{\kappa}
\newcommand{\equ}[1]{\begin{equation} #1 \end{equation}}
\newcommand{\ali}[1]{\begin{align} #1 \end{align}}
\newcommand{\eref}[1]{eq.~(\ref{#1})}
\newcommand{\fref}[1]{fig.~\ref{#1}}
\newcommand{\ddotp}[1]{\frac{d^d #1}{(2\pi)^d}}	
\newcommand{\nnnl}{\nonumber\\}	
\newcommand{\G}[1]{\Gamma(#1)}
\newcommand{\nq}{\nu_1}	
\newcommand{\nw}{\nu_2}	
\newcommand{\nd}{\nu_3}	
\newcommand{\de}{\delta} 
\newcommand{\fig}[4]{\begin{figure}[#1]\centering\epsfig{file=#3}\caption{#4}\label{#2}\end{figure}}

\allowdisplaybreaks 

\title{On the infrared scaling solution of $SU(N)$ Yang-Mills theories\\in the maximally Abelian gauge}

\author{M. Q. Huber}
\email{markus.huber@uni-jena.de}
\affiliation{Institut f\"ur Physik, Karl-Franzens-Universit\"at Graz, Universit\"atsplatz 5, 8010 Graz, Austria}

\author{K. Schwenzer}
\email{schwenzer@physics.wustl.edu}
\affiliation{Department of Physics, Washington University, St. Louis, MO 63130, USA}

\author{R. Alkofer}
\email{reinhard.alkofer@uni-graz.at}
\affiliation{Institut f\"ur Physik, Karl-Franzens-Universit\"at Graz, Universit\"atsplatz 5, 8010 Graz, Austria}

\date{\today}

\begin{abstract}
\noindent 
An improved method for extracting infrared exponents from functional equations is presented. The generalizations introduced allow for an analysis of quite complicated systems such as Yang-Mills theory in the maximally Abelian gauge. 
Assuming the absence of cancellations in the appropriately renormalized integrals the only consistent scaling solution yields an infrared enhanced diagonal gluon propagator in support of the Abelian dominance hypothesis.
This is explicitly shown for $SU(2)$ and subsequently verified for $SU(N)$, where additional interactions exist. We also derive the most infrared divergent scaling solution possible for vertex functions in terms of the propagators' infrared exponents. We provide general conditions for the existence of a scaling solution for a given system and comment on the cases of linear covariant gauges and ghost anti-ghost symmetric gauges.
\end{abstract}


\maketitle

\section{Introduction}\label{sec:intro}

Functional approaches like Dyson-Schwinger equations (DSEs), see e.~g. \cite{Alkofer:2000wg,Fischer:2006ub}, or renormalization group equations (RGEs) \cite{Berges:2000ew,Pawlowski:2005xe} have become standard tools to investigate the infrared (IR) behavior of Landau gauge Yang-Mills theory \cite{vonSmekal:1997vx,vonSmekal:1997is,Pawlowski:2003hq}
during the last few years. The use of stochastic quantization \cite{Zwanziger:2001kw,Zwanziger:2002ia,Zwanziger:2003cf} leads to equivalent results, especially the same IR behavior is found for a power law type solution in the deep IR.  The IR enhanced quantities that lead to confinement of gluons and quarks have been identified as the ghost propagator \cite{vonSmekal:1997is,Lerche:2002ep,Zwanziger:2001kw,Huber:2009tx} and in turn the quark-gluon vertex \cite{Alkofer:2006gz,Alkofer:2008tt}. There are several reasons why Landau gauge was the first choice for such methods: First of all the ghost-gluon vertex does not acquire an IR enhanced dressing as suggested by Taylor \cite{Taylor:1971ff} and confirmed later on by DSE calculations \cite{Schleifenbaum:2004id,Alkofer:2008jy,Fischer:2006vf,Fischer:2009tn} and lattice simulations \cite{Cucchieri:2004sq,Ilgenfritz:2006he,Cucchieri:2008qm}. This provided for a long time the starting point of the IR analysis. Second, Landau gauge is the simplest gauge in terms of DSEs, since it has only two fields and three interactions, which are the lowest numbers possible for a local gauge condition. Other gauges such as Coulomb gauge, the maximally Abelian gauge (MAG) or ghost anti-ghost symmetric gauges involve more degrees of freedom and more interactions rendering the functional equations more intricate.

Numerical solutions for the coupled system of integral equations of the ghost and gluon propagators were first obtained in ref. \cite{vonSmekal:1997is} and steadily developed by including quarks \cite{Fischer:2003rp} or improving the truncations to get better transversality properties of the gluon propagator \cite{Fischer:2008uz}. The quark-gluon vertex and the quark propagator have been investigated as a coupled system \cite{Alkofer:2008tt}, whereas the other three-point functions \cite{Schleifenbaum:2004id,Alkofer:2008dt} and the four-gluon vertex \cite{Kellermann:2008iw} have been treated in a semi-perturbative approach. Again this is only possible because of the ghost-gluon vertex, which has a constant dressing function in the IR, and the dominant role of the ghosts. 
Having first results a posteriori enabled the investigation of the IR regime in a more general way. 
For not primitively divergent vertex functions a qualitative solution can be derived in $d$ dimensions in terms of power laws for the dressing functions \cite{Alkofer:2004it,Huber:2007kc,Alkofer:2007hc}.
The combination of DSEs and RGEs provides the means to identify a unique IR scaling solution \cite{Fischer:2006vf,Fischer:2009tn}, as does the assumption of a stable skeleton expansion \cite{Alkofer:2008jy}. Indeed, as we will show below, the latter is a consequence of constraints derived from RGEs.
In the light of the success of a scaling analysis of the low momentum behavior of Yang-Mills theory in Landau gauge, it seems promising to pursue this method also in other gauges like the MAG.

In general DSEs and RGEs allow two different types of solutions in Landau gauge \cite{Fischer:2008uz,Alkofer:2008jy}, called the scaling solution, in which the dressing functions are related via scaling relations \cite{vonSmekal:1997is,Lerche:2002ep,Zwanziger:2001kw}, and the decoupling solution \cite{Boucaud:2008ji,Aguilar:2008xm,Fischer:2008uz,Dudal:2007cw,Alkofer:2008jy}. The latter features a constant gluon propagator at zero momentum (sometimes interpreted as a gluon screening mass) and a tree-level like ghost propagator. Both solutions can be obtained within the framework of DSEs and RGEs, depending on the choice of boundary conditions  \cite{Fischer:2008uz}, and it has been shown in ref. \cite{Alkofer:2008jy} that only the scaling solution leads to IR enhanced vertex functions.
However, in order to have static quark confinement general bounds on the IR exponents of the ghost and gluon propagators must be satisfied \cite{Braun:2007bx}. Both solutions fulfill this criterion.
Another issue of the decoupling solution is the breaking of global gauge and BRST symmetries as discussed in ref. \cite{Fischer:2008uz}. Lattice simulations \cite{Cucchieri:2007md,Cucchieri:2007rg,Cucchieri:2008fc,Bogolubsky:2009dc,Bornyakov:2008yx,Pawlowski:2009iv} and the refined Gribov-Zwanziger framework \cite{Dudal:2007cw,Dudal:2008sp} seem to favor the decoupling solution, but for the former there are still unresolved issues concerning discretization \cite{Sternbeck:2008na} and gauge fixing \cite{Maas:2008ri}.\footnote{Recently an interesting suggestion was made how to implement different choices of the boundary conditions similar to functional equations also on the lattice \cite{Maas:2009se}. By selecting the Gribov copies with respect to the value of the ghost dressing function, different behaviors for the ghost propagator were found.}

Whereas the Landau gauge has been used intensively for the last ten years in the context of non-perturbative investigations, the MAG was less common. The reason might be its intrinsic complexity compared to the Landau gauge. However, it offers an appealing picture of confinement, in which chromoelectric charges are confined by the condensation of chromomagnetic monopoles \cite{Mandelstam:1974pi,'tHooft:1976ac}. Thereby the chromoelectric flux is squeezed into flux tubes by the Meissner effect. At the ends of these tubes the chromoelectric charges are located. This confinement scenario works in analogy to the conventional superconductor, but with the role of electric and magnetic charges exchanged. Thus it is called a dual superconductor. In compact $U(1)$ lattice gauge theory it is known that at low values of $\beta$, i.~e. for large coupling, magnetic monopoles indeed do condense and confine electric charges \cite{'tHooft:1981ht,DeGrand:1980eq}. One can try and generalize this to non-Abelian gauge theories by a specific choice of gauge, so that the theory can be expressed as an Abelian theory with magnetic monopoles \cite{'tHooft:1981ht}. A special choice of this class of gauges is the MAG, where the off-diagonal components of the gauge-field are gauge-fixed such that their norm is minimized with respect to gauge transformations \cite{'tHooft:1981ht}. The chromomagnetic monopoles then can confine all chromoelectric charges if they condense. The appearance of these topological objects was first investigated by lattice calculations in refs. \cite{Kronfeld:1987ri,Kronfeld:1987vd}.

The existence of monopoles and their role for confinement was the topic of a recent lattice calculation, where no gauge was fixed \cite{Suzuki:2009xy}. The string tension calculated from the full gauge configurations, the Abelian part and the monopole part showed an amazing agreement. This hints at the importance of the Abelian part for confinement. Since monopoles are gauge dependent objects, it is not completely clear what role they play in different gauges, as for example they don't exist in Landau gauge. But it seems that the MAG is very well suited to identify the physical contributions of monopoles. Since it is easier to extract monopoles in the MAG than in gauge unfixed calculations like in ref. \cite{Suzuki:2009xy}, it can provide a useful means of extracting the physical part of monopoles on the lattice \cite{Suzuki:2009xy}. On the other hand gauge fixing is necessary for all functional approaches. Therefore the MAG may be the best way to investigate monopoles with these methods. However, to establish a connection to monopoles is beyond the scope of this article.

The dual superconductor picture of confinement leads to the hypothesis of Abelian dominance \cite{Ezawa:1982bf}: Since the classical configurations of the monopoles live in the Cartan subalgebra, i.~e. the maximal Abelian subalgebra, of the gauge group, the relevant degrees of freedom should be Abelian.
This can be interpreted as a mass for the off-diagonal gluons \cite{Amemiya:1998jz}, so that they decouple below this momentum scale and do not influence the dynamics in the IR. Indeed lattice results \cite{Amemiya:1998jz,Bornyakov:2003ee,Mendes:2008ux} indicate that the off-diagonal gluon propagator is suppressed at low momenta compared to the diagonal propagator. An analytic study did also find a mass for the off-diagonal gluon \cite{Dudal:2004rx}. This result holds, if one takes into account the existence of the Gribov horizon (Gribov-Zwanziger framework) \cite{Capri:2005tj,Capri:2006cz}. The diagonal gluon propagator then shows a Gribov-type behavior, i.~e. it vanishes at zero momentum. However, by introducing additional terms into the Lagrangian that correspond to non-vanishing condensates (refined Gribov-Zwanziger framework) all three propagators of the MAG were found to possess a massive IR behavior \cite{Capri:2008ak}.

In this article we mention another possible manifestation of Abelian dominance: The diagonal gluon propagator is enhanced in the IR similar to the ghost propagator in Landau gauge. Therefore at low momenta it constitutes the dominant degree of freedom, which determines the behavior of the other two propagators and all vertices. However, since our analysis can only be the first step towards a full solution of the propagators in the MAG, we only mention this possibility and leave further discussions for the future, when more detailed results are available.

There have also been investigations of the MAG on the lattice \cite{Amemiya:1998jz,Bornyakov:2003ee,Mendes:2008ux}. The data for the lowest reached momenta showed that all three propagators are IR finite \cite{Mendes:2008ux}. This agrees with results from the refined Gribov-Zwanziger framework \cite{Capri:2008ak}, where the addition of condensates also leads to the same IR properties. However, there has not been a successful attempt to use dynamical functional approaches to assess the low momentum behavior yet.\footnote{An earlier attempt to investigate the MAG with DSEs was performed in ref. \cite{Shinohara:2003mx}, where all two-loop contributions were neglected and the issue of bare two-point functions in the DSEs was not addressed, see sec. \ref{sec:General-Approach}. In this article we will show that under these circumstances no consistent scaling solution can emerge.} The question is if there is a similar discrepancy between different methods as in the Landau gauge and if yes, what qualitative properties a scaling solution in the MAG might have. To fully answer this requires a complete numerical solution of the DSEs (or other functional equations) at all scales. For such an endeavor knowledge about the asymptotic behavior of Green functions proves useful \cite{vonSmekal:1997vx,Maas:2005xh}.
For high momenta we can rely on perturbative calculations, which have been performed up to three loops in the MAG \cite{Gracey:2005vu}. Since perturbation theory is an expansion in the coupling constant it is clear how to classify contributions by the number of loops and the UV leading part consists of one-loop diagrams.
For asymptotically low momenta, however, no information is available. In this article we will attempt such an analysis, where one interesting result will be that the structure of IR dominant diagrams is quite different from all other known examples: Instead of one-loop diagrams (as in the Landau gauge) two-loop diagrams are IR leading. This constitutes the first known case of this type and prohibits the use of known approximations for a numerical solution.

The investigations of the MAG mentioned above were all done for the gauge group $SU(2)$. It is an interesting feature of this gauge that the number of interaction terms in the Lagrangian is different for $SU(2)$ and higher $SU(N)$. Thus here the question of the gauge group dependence is of special importance, in contrast to Landau gauge, where both functional approaches and lattice simulations \cite{Sternbeck:2007ug, Cucchieri:2007zm,Maas:2007af} agree that the qualitative behavior is the same for $SU(2)$ and $SU(3)$. Since the DSEs in the MAG are even more complex for $SU(3)$ than for $SU(2)$ the need for a more effective method to investigate the IR regime is obvious. The scheme presented below provides this and shows that the additional interactions of $SU(3)$ neither alter the solution derived in $SU(2)$ nor allow for an additional one.

The article is structured as follows: In sec. \ref{sec:Power-Counting} we illustrate the basic idea of the method of power counting to determine qualitative IR scaling solutions. It serves as an entry point for the reader new to this kind of analysis and may be skipped by the reader familiar with this topic. Sec. \ref{sec:General-Approach} explains the method we will employ to investigate the MAG. It is not specific to the MAG and thus the results may be used for other gauges as well. However, we will repeatedly exemplify some steps in the Landau gauge to provide a connection from the abstract method to familiar facts and relations. The MAG and other gauges are investigated in sec. \ref{sec:IR-Sol-MAG} and sec. \ref{sec:Other-Gauges}, respectively. Our conclusions are presented in sec. \ref{sec:Conclusions}.
Some detailed calculations are deferred to three appendices.

\section{The Infrared Power Counting Technique}
\label{sec:Power-Counting}

In our study of the IR behavior of Yang-Mills theory the quantities of interest are the full non-perturbative Green functions. Usually one starts the analysis by considering two-point functions, i.~e. propagators.
They are parameterized by
\begin{align}
\Delta_{ij}(p)=P_{ij}\frac{Z(p^2)}{p^2},
\end{align}
where $P_{ij}$ is the part containing color and Lorentz structure  and $Z$ is some dressing function. For low momenta, i.~e. $p^2\rightarrow 0$, the latter obeys due to renormalization group arguments a power law,
\begin{align}
Z^{IR}(p^2)=A\cdot(p^2)^\alpha,
\end{align}
with some constant $A$ and the quantity $\alpha$ called infrared exponent (IRE). Clearly if $\alpha<0$ the propagator is enhanced in the IR compared to the tree-level propagator and it vanishes for $\alpha>1/2$. We assumed here for simplicity that the propagator can be described with a single dressing function, but a generalization to several dressing functions is possible. Although vertices are more complicated and feature in general many dressing functions corresponding to different tensors, all of them show the same momentum dependence when all momenta vanish. This case is in the following denoted as the \textit{uniform} limit and will be studied in this work.

Although DSEs can be used to determine dressing functions for the complete momentum region, we are only interested here in the behavior at low external momenta of the integrals. Since they contain terms like $1/(p-k)^2$, where one momentum is the loop momentum and the other the external one, main contributions to the integrals arise from loop momenta that have the same order of magnitude as the external ones. When all momenta are small, we can replace the dressing functions by their respective IR expressions. Due to dimensional reasons all momenta of the integral transform to external momenta upon integration. For two-point functions this can directly be seen from the available analytical expression for the one-loop integral
\footnote{From UV and IR convergence one can see that the exponents $\nu_1$ and $\nu_2$ have to fulfill $d/2+\nu_1+\nu_2 \leq0$, $d/2+\nu_1\geq0$ and $d/2+\nu_2\geq0$.}
\begin{align}\label{eq:2-point}
\int\frac{d^{d}k}{\left(2\pi\right)^{d}}\left(k^{2}\right)^{\nu_{1}}\left(\left(k-p\right)^{2}\right)^{\nu_{2}}=\left(4\pi\right)^{-\frac{d}{2}}\frac{\Gamma\left(\frac{d}{2}+\nu_{1}\right)\Gamma\left(\frac{d}{2}+\nu_{2}\right)\Gamma\left(-\frac{d}{2}-\nu_{1}-\nu_{2}\right)}{\Gamma\left(-\nu_{1}\right)\Gamma\left(-\nu_{2}\right)\Gamma\left(d+\nu_{1}+\nu_{2}\right)}\left(p^{2}\right)^{\frac{d}{2}+\nu_{1}+\nu_{2}} .
\end{align}
Similar expressions can be obtained for the two-loop integrals of two-point functions by successive integration of the integrals. In case of higher vertex functions analytic results are only known for one-loop three-point functions \cite{Davydychev:1991va,Anastasiou:1999ui}. An explicit calculation of three-point functions in Landau gauge has been presented in refs. \cite{Huber:2008mq,Alkofer:2008dt}. A very successful method for the evaluation of these kind of integrals is the negative dimensions integration method (NDIM) \cite{Dunne:1987am,Halliday:1987an,Dunne:1987qb,Ricotta:1990nd}.
In this work we only treat the case of uniform IR divergences, i.~e. all external momenta tend to zero simultaneously. The case of only a subset of the external momenta going to zero is discussed in refs. \cite{Alkofer:2008jy,Alkofer:2008dt}. For an explanation of how the power counting works for RGEs see ref. \cite{Fischer:2006vf}.

Knowing that all integrals transform into powers of the external momenta one can work directly at the level of the IREs (but has then to ignore possible cancellations of integrals). We want to make this point clear with a simple example. Consider the gluon propagator DSE in Landau gauge, which is given diagrammatically in \fref{fig:LG-g-DSE}.
\fig{th}{fig:LG-g-DSE}{gluon-DSE,width=0.7\linewidth}{The gluon propagator DSE in Landau gauge: The bare propagator, the ghost-loop, the gluon-loop, the tadpole, the squint and the sunset diagrams. Small dots represent bare vertices and large dots 1PI vertices.}
Evaluation of integrals yields
\ali{
A^{-1} (p^2)^{1-\delta_{gl}} = p^2-(p^2)^{1+\delta_{3g}+2\delta_{gl}} L^{g-loop}+(p^2)^{1+\delta_{gg}+2\delta_{gh}} L^{gh-loop} - (p^2)^{1+\delta_{4g}+3\delta_{gl}} L^{sunset} - (p^2)^{1+2\delta_{3g}+4\delta_{gl}} L^{squint},
}
with the IREs $\delta_{gl}$, $\delta_{gh}$, $\delta_{3g}$, $\delta_{gg}$ and $\delta_{4g}$ of the gluon and ghost propagators, the three-gluon, the ghost-gluon and the four-gluon vertices. The $L$s denote constant terms for the gluon-loop, the ghost-loop, the sunset and the squint diagrams.\footnote{The tadpole was not taken into account as it does not depend on the external momentum and can thus be absorbed in the renormalization.} They depend on the IREs (see \eref{eq:2-point} above) and contain also the constant coefficients from the power laws. When $p$ is small not necessarily all terms on the right-hand side of the equation scale equally, but at least one must have the same exponent as the left-side, what can be phrased as
\ali{\label{eq:max}
(p^2)^{-\delta_{gl}} = \max\left((p^2)^0,(p^2)^{\delta_{3g}+2\delta_{gl}},(p^2)^{\delta_{gg}+2\delta_{gh}} , (p^2)^{\delta_{4g}+3\delta_{gl}}, (p^2)^{2\delta_{3g}+4\delta{g}}\right).
}
Canonical dimensions do not occur here any longer, since they cancel in four dimensions. Eq. (\ref{eq:max}) is equivalent to
\ali{\label{eq:min}
-\delta_{gl} = \min(0,\delta_{3g}+2\delta_{gl},\delta_{gg}+2\delta_{gh}, \delta_{4g}+3\delta_{gl}, 2\delta_{3g}+4\delta{g}).
}
From this equation we can extract single inequalities, e.~g.
\ali{
0 \leq \delta_{3g}+3\delta_{gl},
}
which can be used to show that $\de_{gl}\geq0$: Since the three-gluon vertex is primitively divergent its DSE features the bare vertex, so that we have $\de_{3g}\leq0$, i.~e. the vertex' IRE is non-positive. Taking this into account we get
\ali{
0\leq -\de_{3g} \leq \de_{gl} \quad \Rightarrow \quad \de_{gl}\geq 0.
}
In sec. \ref{sec:General-Approach} we will show how to single out the relevant inequalities, so that one does not have to work with inconvenient minimum functions.

First IR analyses of DSEs and RGEs in Landau gauge relied on investigating individual diagrams. Whereas at first only the IR behavior of the propagators and the ghost-gluon vertex was determined \cite{vonSmekal:1997vx,Lerche:2002ep,Zwanziger:2001kw}, this could be extended later on to the whole tower of Green functions by use of a so-called skeleton expansion \cite{Alkofer:2004it}. Thereby higher n-point functions are expanded in loops consisting of dressed primitively divergent vertex functions. Another approach was taken in \cite{Fischer:2006vf,Fischer:2009tn}, where the combination of the two equivalent systems of DSEs and RGEs allowed to show the uniqueness of the solution for the IREs. In \cite{Alkofer:2008jy} it was shown that also the assumption of a non-explicitly divergent skeleton expansion allows a proof of uniqueness. Non-explicitly divergent means here that insertions, necessary to get higher orders of the expansion, do not have negative IREs. Since, as shown below, the constraints obtained from this assumption follow also from the RGEs without such an assumption, this result is not surprising. Furthermore in \cite{Alkofer:2008jy} the existence of additional IREs for certain kinematic limits, e.~g. only one momentum of a three-point function going to zero, was also established.
A numerical confirmation of kinematic IR singularities for the three-point functions of Landau gauge Yang-Mills theory has been presented in ref. \cite{Alkofer:2008dt}, where it was also shown that they only appear in the longitudinal parts of these vertices. 
Also the quark-gluon vertex features kinematic singularities \cite{Alkofer:2008tt,Schwenzer:2008vt}, which turn out to be as strong as the uniform ones. This is in contrast to the pure Yang-Mills case, where they are significantly weaker \cite{Alkofer:2008dt,Fischer:2009tn}.

The first step towards the improved technique used in this article can be found in \cite{Huber:2007kc}, where a formula for the IREs of all diagrams in a DSE was presented. Working in Landau gauge the known exponents of propagators and the ghost-gluon vertex served as input. However, that result still was based on the use of a skeleton expansion, i.~e. only three- and four-gluon vertices and no other vertex functions were taken into account. A generalization of this formula to arbitrary interactions and without assumptions about any of them is used in the next section to derive a lower bound for vertex IREs.

\section{Maximally infrared divergent solution and scaling relations}\label{sec:General-Approach}

In this section we give an overview of the method we are going to employ in sec. \ref{sec:IR-Sol-MAG} to derive an IR solution for the MAG. We specialize on pure scaling solutions and do not consider dynamical mass generation or kinematic divergences. The analysis is similar to the one presented in the recent ref. \cite{Fischer:2009tn} for Landau gauge.  We will keep the discussion general without fixing the number or type of fields yet, since doing so does not provide any simplification; on the contrary, writing some equations explicitly for the MAG leads to unnecessarily long expressions.
The crucial results of this section might prove useful also in the investigation of other actions, as we identify the most restrictive inequalities one can derive from the system of functional equations in the pure scaling case. Up to now the systems of equations were studied diagram by diagram, whereas here we provide a more general approach that enables one to write down the decisive inequalities.
These contain the only information as extracted from functional equations that restrict the IR behavior. All other inequalities, of which there are infinitely many, arising from individual diagrams are superfluous in any investigation and can be disregarded. 
As we do not employ any truncation, we analyze the \textit{full system of DSEs and RGEs}.
In principle it would not even be necessary to write down the equations explicitly, since the necessary information is encoded in the interaction types of the action.
Thus with the results of this section one can then get the IR scaling solution rather easily only by considering the type of interactions in the Lagrangian without explicit power counting of individual graphs.

Although the results are simple, the derivation is somewhat technical. Below we try to explain the idea and defer some details to appendices. To allow a more intuitive understanding of some relations we give examples in the Landau gauge, since many readers might be familiar with it. We start with a general formula for the IRE of an arbitrary diagram derived purely by combinatorics. It will lead to the notion of the maximally IR divergent solution. Then we analyze the propagator equations and show that at least one primitively divergent vertex is not IR enhanced.

For the existence of a scaling solution we have to make some assumptions:
\begin{itemize}
 \item An IR enhanced propagator, i.~e. $\de<0$, is only possible if the term arising from the bare propagator is canceled in the DSE.
 This is simply a consequence of the general form of propagator DSEs, where the inverse dressed propagator is on the left-hand side and the inverse bare propagator on the right hand side, yielding the equation
\ali{
-\de=min(0,\ldots) \quad \Rightarrow \quad \de\geq 0.
}
Thus the bare propagator has to be canceled somehow to allow for IR enhancement. This is realized in Landau gauge for the ghost propagator by the so-called horizon condition \cite{Zwanziger:1992qr}, which is implemented into the DSE of the ghost propagator by an appropriate renormalization condition. We work here under the assumption that similar mechanisms exist in other gauges, so that bare propagators can vanish effectively from the DSE at zero momentum. We comment on the possibility $\de=0$ below at the end of subsec. \ref{ssec:Prop-Eqs}.

\item Propagators with several dressing functions lead to several DSEs that are obtained by projecting out the corresponding tensor parts from the original DSE. Thus the individual parts of the propagators can be taken into account in a power counting analysis by different IREs. The vertices are split accordingly. However, different IREs are only expected if the structure of the inequalities changes. This could happen for example if the bare vertex vanishes in the projection of a vertex equation. On the other hand, an IRE can also be determined by constraints like symmetries. This happens for example in linear covariant gauges \cite{Alkofer:2000wg}, where the gauge symmetry determines the longitudinal part of the gluon propagator (see sec. \ref{sec:Other-Gauges}). This leads to additional information that can be used in the analysis.

 \item Cancellations always are a possible threat for a scaling analysis if terms are taken into account that actually disappear. We disregard this possibility here and assume that none of the important contributions behaves in this way. Indeed the analysis is only rendered invalid if the IR leading terms vanish, while a vanishing of subleading terms is not relevant.
\end{itemize}

\subsection{A formula for the infrared exponent of an arbitrary diagram}
\label{ssec:formula-IRE}

Before we derive any formulae we explain our notation. We do not specify any fields explicitly in our derivations for two reasons: First, it is not easy to write down the derivations in the MAG, because this gauge has so many interactions between its three fields that the equations become quite large. In our generic notation the derivation can be done in a compact way. Second, proceeding in a general way we can learn more about the basic structure of IR scaling solutions and the results can be used for various actions. So  we denote the fields by the set $\{\phi_s\}$, $s=1, \ldots, S$, with $S$ the number of different fields in the action. In the MAG this is $\{A,B,\bar{c},c\}$ (see sec. \ref{sec:IR-Sol-MAG} for details) and in Landau gauge $\{A,\bar{c},c\}$. Furthermore we need the following quantities:
\begin{itemize}
\item $\delta_{i}$: The IRE of the propagator of the field $\phi_i$.
 \item $\delta_{i\ldots k}$: The IRE of a vertex with legs of the fields $\phi_i$, $\ldots$, $\phi_k$.
 \item $n_{i\ldots k}$: The number of vertices of the type $\phi_i \ldots \phi_k$. Superscripts $d$ or $b$ denote only dressed or bare vertices, respectively, whereas no superscript refers to both.
 \item $n_{i}$: The number of internal propagators of the field $\phi_i$.
 \item $m_{i}$: The number of external legs of the field $\phi_i$.
 \item $c_{i\ldots j}$: The canonical dimension of the vertex indicated by the subscript.
 \item $k_{{i}}^{j\ldots l}$: The number of legs of the field $\phi_i$ the vertex $\phi_j \ldots \phi_l$ has.
\end{itemize}

One of the main pieces in the forthcoming analysis will be a formula for the IRE $\de_v$ of an arbitrary diagram $v$ with canonical dimension $c_v$. With the notation introduced above we can write down this general formula for an arbitrary $l$-loop diagram in $d$ dimensions:
\begin{align}
\delta_{v}= & l\frac{d}{2}+\sum_{i}n_i(\delta_{{i}}-1)+\sum_{vertices,r\geq3}n^{d}_{{i_1}\ldots {i_r}}(\delta_{{i_1}\ldots {i_r}}+c_{{i_1}\ldots {i_r}})+\\
 & +\sum_{vertices,r\geq3}n^{b}_{{i_1}\ldots {i_r}}c_{{i_1}\ldots {i_r}}-c_{v}.
\end{align}
The double subscripts appearing in the last two sums indicate all possible combinations of $r$ fields.
E.~g. for the Landau gauge the term corresponding to $r=3$ is
\ali{
n^d_{AAA}\left(\de_{AAA}+\frac1{2}\right)+n^d_{A\bar{c}c}\left(\de_{A\bar{c}c}+\frac1{2}\right).
}
Using topological relations this expression is reformulated such that the dependence on internal propagators is exchanged for one on the number of external legs. The details of this calculation are given in Appendix \ref{sec:IRE-Formula}. The final result in four dimensions is
\begin{align}\label{eq:master-formula}
\setlength{\fboxsep}{3mm}
\fbox{$ \displaystyle \delta_v= -\frac{1}{2}\sum_{i}m_i\delta_{{i}}+\sum_{vertices,r\geq3}n^{b}_{{i_1}\ldots {i_r}}\left(\frac{1}{2}\sum_{i}k_{{i}}^{{i_1}\ldots {i_r}}\delta_{{i}}\right)
  +\sum_{vertices,r\geq3}n^{d}_{{i_1}\ldots  {i_r}}\left(\delta_{{i_1}\ldots {i_r}}+\frac{1}{2}\sum_{i}k_{{i}}^{{i_1}\ldots {i_r}}\delta_{{i}}\right).$}
\end{align}

The sums extend over all vertices with $r$ legs in the diagram. Eq. (\ref{eq:master-formula}) is a purely combinatoric result without any assumptions whatsoever employed.
At this point one may be tempted to employ a skeleton expansion, i.~e. restrict the sums to primitively divergent vertices. This is indeed possible since the derivation of \eref{eq:master-formula} was valid for a generic diagram, i.~e. also those one gets in a skeleton expansion. Nevertheless an analysis without skeleton expansion can be done and we do not have to rely on the assumption of a stable skeleton expansion here. Note that for RGEs $n^b=0$ and for every diagram in a DSE there is only one $n^b=1$, while the other $n^b$ are 0. For a consistent solution of DSEs and RGEs \eref{eq:master-formula} hints already at the condition that at least one coefficient of an $n^b$ is zero, leading in turn to a non-scaling vertex. This consistency condition was first stated in ref. \cite{Fischer:2006vf}. Below we will show in subsec. \ref{ssec:Prop-Eqs} in detail how it emerges for general structures of interactions.

\begin{figure}[th]
\begin{minipage}{0.48\textwidth}
\epsfig{file=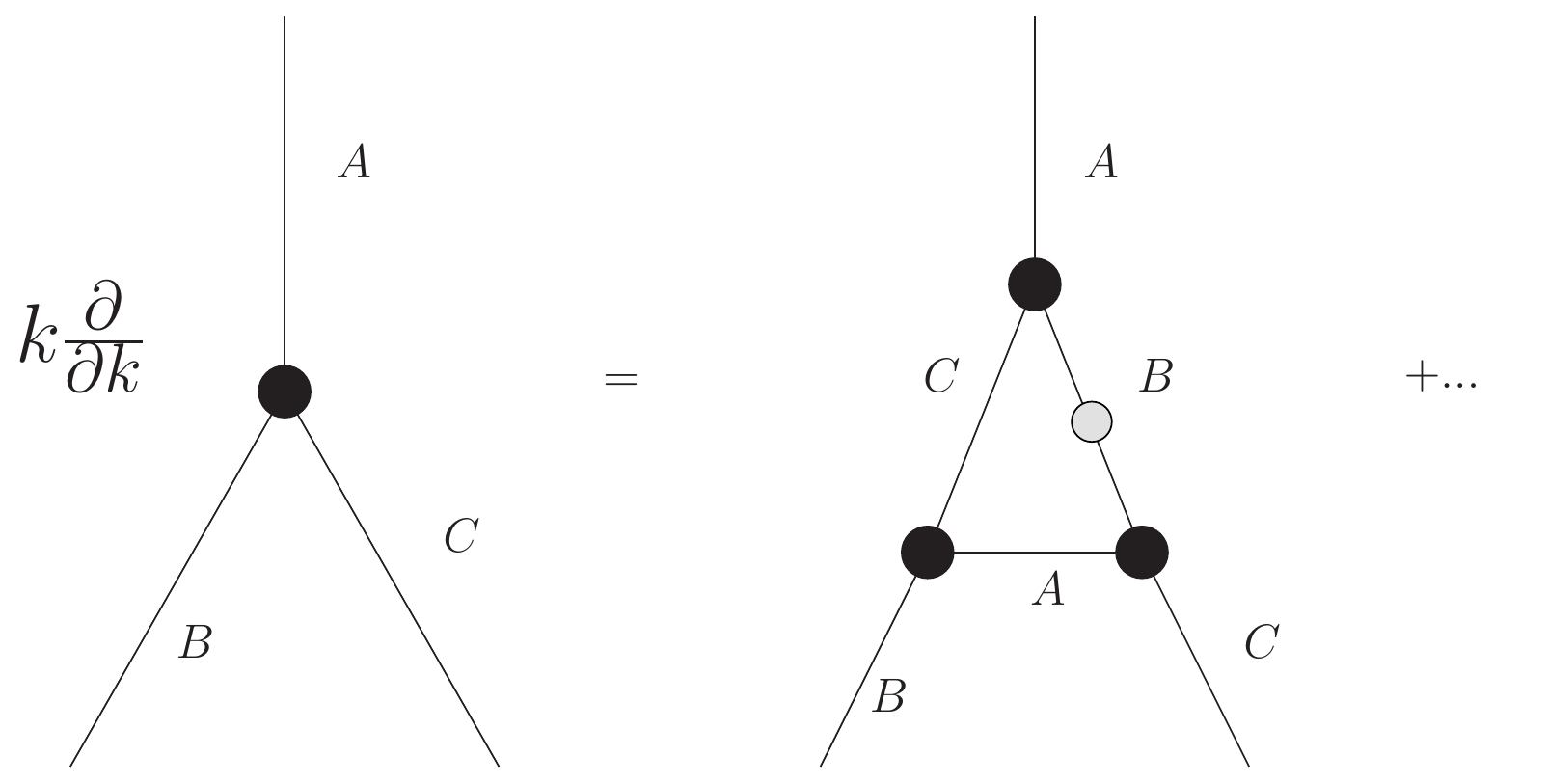,width=1\textwidth}
\caption{\label{fig:RG-3-point}One specific diagram in the RGE of a generic three-point function. Internal lines represent dressed propagators, black blobs dressed vertices. The gray blob is a regulator.}
\end{minipage}
\end{figure}

Eq. (\ref{eq:master-formula}) leads to the notion of a maximally IR divergent solution. Therefor we need information about the coefficients of $n^d$, which we can get from RGEs, and $n^b$ in the sums. We explain the case of three-point functions and shift the derivation for higher vertex functions to Appendix \ref{sec:Constraints}.
Consider the RGE for a generic three-point function with the fields $A$, $B$ and $C$, 
where it is not excluded that some of them are the same. For instance $A$ may be a gluon field and $B$ and $C$ the Faddeev-Popov ghost and anti-ghost fields of Landau gauge.
In \fref{fig:RG-3-point} we only show the one diagram on the right-hand side we are interested in. At the level of IREs we have on the left-hand side only the IRE of the three-point function; on the right-hand side appears three times the same three-point function and one IRE for each propagator. Since the diagram on the right-hand side cannot be more IR divergent than the left-hand side, we can write down an inequality for the IREs:
\begin{align}\label{eq:RG-3-point}
\delta_{ABC}\leq 3\delta_{ABC}+\delta_A+\delta_B+\delta_C \qquad \Rightarrow \qquad \delta_{ABC}+\mhalfo(\delta_A+\delta_B+\delta_C)\geq0.
\end{align}
Note that in four dimensions for uniform scaling the canonical dimensions always cancel and need not be considered explicitly, but the argument is also valid in $d$ dimensions; for details see Appendix \ref{sec:Constraints}. Eq. (\ref{eq:RG-3-point}) implies that the coefficient of $n^d_{ABC}$ in \eref{eq:master-formula} is non-negative for arbitrary fields $A$, $B$ and $C$. If we can find other similar constraints for higher vertex functions, we know that the sum over dressed vertices can only increase but never lower the IRE of a diagram. Indeed such inequalities exist and their derivation can be found in Appendix \ref{sec:Constraints}. The general form is
\begin{align}\label{eq:verts-props-inequal1}
\setlength{\fboxsep}{3mm}
\fbox{$ \displaystyle \delta_{{i_1}\ldots {i_r}}+\frac{1}{2}\sum_{i}k_{{i}}^{{i_1}\ldots {i_r}}\delta_{{i}} \geq 0.$}
\end{align}

Interestingly it turns out that for primitively divergent vertex functions the same restrictive inequalities arise from the assumption of a non-divergent skeleton expansion or, phrased differently, the skeleton expansion cannot lead to more divergent terms. Such an expansion expresses non-primitively divergent vertex functions by a loop expansion containing only dressed quantities. Higher orders can be obtained by inserting additional loops using basic insertions, which consist only of primitively divergent vertex functions, into a diagram \cite{Alkofer:2004it}. By non-divergence we mean that the IRE of such an insertion is greater or equal to zero so that the new diagram is not more IR divergent than the original one. Using the skeleton expansion one gets a series of infinitely many diagrams and it is a priori not clear, whether the sum can change the IR scaling. The general result from above shows that this is actually \textit{not} the case and previous results based on the skeleton expansion are indeed valid.
Note that at least in Landau gauge it was found for two, three and four dimensions that the insertions do not change the IREs, i.~e. the inequalities above are saturated \cite{Alkofer:2004it,Huber:2007kc}.

The inequalities above also restrict the coefficients of $n^b$ in the second sum of \eref{eq:master-formula}. 
We know that for primitively divergent vertex functions the bare vertex appears on the right-hand side of its DSE. Therefore these n-point functions have an IRE lesser than or equal to zero,
\begin{align}\label{eq:prim-div}
\delta_{prim.\,div.\,vertices}\leq0,
\end{align}
and not only the coefficients of the $n^d$ are non-negative, but also those of the $n^b$, since they only differ by a non-positive number. The fact that \eref{eq:prim-div} is only valid for primitively divergent vertex functions causes no problem, since $n^b\neq0$ only for primitively divergent vertex functions.
We demonstrate explicitly how this argument works for a three-point function:
\begin{align}\label{eq:RG-3-point-bare}
-\frac{1}{2}(\delta_{A}+\delta_{B}+\delta_{C})\leq\delta_{ABC}\leq0 \qquad \Rightarrow \qquad \delta_{A}+\delta_{B}+\delta_{C}\geq0.
\end{align}
For a general primitively divergent interaction $\phi_{i_1} \ldots \phi_{i_r}$ this can be written as
\begin{align}\label{eq:prop-IRE-inequal1}
\setlength{\fboxsep}{3mm}
\fbox{$ \displaystyle \frac1{2} \sum_{i} k_{i}^{{i_1} \ldots {i_r}} \de_{i} \geq 0.$}
\end{align}
This inequality can also be derived with the same argument from DSEs, since DSEs yield similar inequalities as those in \eref{eq:verts-props-inequal1}, but with a different numerical factor in front of the vertex IRE.

Thus the second and third terms in \eref{eq:master-formula} are both non-negative and the \textit{maximally IR divergent solution}, i.~e. that with the lowest IRE possible, for n-point functions with $m_{i}$ legs of the field $\phi_i$ in terms of the propagator IREs $\delta_{i}$ is
\begin{align}\label{eq:Maximal-IR-Divergent}
\setlength{\fboxsep}{3mm}
\fbox{$ \displaystyle \delta_{v,max}= -\frac{1}{2}\sum_{i}m_{i}\delta_{{i}}.$}
\end{align}
It is realized when the inequalities connecting the IREs of vertices and propagators, \eref{eq:verts-props-inequal1}, are saturated and thus all coefficients of the $n^d$ vanish. Then all diagrams in an RGE scale equally (since $n^b=0$). Diagrams in DSEs, however, contain a bare vertex, which is reflected by the appearance of $n^b$. Its coefficient vanishes only for the IR leading diagrams, e.g. those with a bare ghost-gluon vertex in Landau gauge, while the other diagrams have larger IREs. In Landau gauge these are the diagrams with bare three-gluon and four-gluon vertices, which have an IRE that is $3\ka$ and $4\ka$ higher than for the leading diagrams, respectively. These numbers stem directly from the coefficients of the $n^b$. Thus for the maximally IR divergent solution the bare vertex divides the diagrams into classes of IR importance.

The important formulae of this subsection are eqs. (\ref{eq:master-formula}), (\ref{eq:verts-props-inequal1}) and (\ref{eq:prop-IRE-inequal1}). The latter two comprise all constraints on the IREs one can get in a uniform scaling analysis.
They can directly be derived from the vertices of the theory and there is no longer the need to write down many inequalities from all sorts of different diagrams. To demonstrate this we again employ the Landau gauge as an example, where one gets up to four-point functions the following inequalities:
\ali{
\de_{gl}\geq&0, & \quad 2\de_{gh}+\de_{gl}\geq&0\nnnl
\de_{4g}+2\de_{gl}\geq&0, & \quad \de_{3g}+\frac3{2}\de_{gl} \geq&0,\nnnl
\de_{gg}+\de_{gh}+\frac1{2}\de_{gl}\geq&0, & \quad \de_{2gh2g}+\de_{gh}+\de_{gl}\geq&0,
}
with $\de_{2gh2g}$ the IRE of the ghost-gluon scattering kernel.
Any attempt to obtain more restrictive inequalities fails.
Thus on the level of the IREs one can \textit{reduce quite large systems of DSEs to a much smaller set of inequalities without making any assumptions or missing any information relevant for the uniform IR scaling}.

\subsection{Analysis of the propagator equations}
\label{ssec:Prop-Eqs}

In this subsection we will have a closer look at the propagator equations. For a given system of DSEs one could of course perform a one-by-one analysis of each diagram. This is a simple task in Landau gauge, where the ghost and gluon propagator equations have one and four diagrams, respectively, neglecting the bare two-point functions and tadpole diagrams. The situation is more involved for the MAG, where in $SU(2)$ we have six, seven and seven diagrams for the diagonal gluon, the off-diagonal gluon and the ghost propagator, respectively. These numbers increase for $SU(3)$ to 13, 17 and 14. Therefore such an analysis easily becomes extensive. We describe below how the analysis can be performed efficiently for all diagrams at once, again by using abstract notation. As a result we can show that for a consistent scaling solution at least one of the inequalities found above has to be saturated and thus at least one vertex cannot be IR enhanced.

In the case that we only have three- and four-point interactions there are only four different types of diagrams besides the bare two-point function: the tadpole, the one-loop, the sunset and the squint diagrams, see \fref{fig:prop-DSE-graphs}. All of these possibilities appear in the DSEs of the MAG in at least one realization, see \fref{fig:MAG-DSEs}. However, the argument given below holds also for higher interactions.
\fig{th}{fig:prop-DSE-graphs}{prop-diagrams,width=\textwidth}{Types of graphs in a two-point DSE: (a) tadpole, (b) one-loop, (c) sunset, (d) squint.}

The analysis is based on the fact that the IRE of the left-hand side has to be matched by at least one IRE on the right-hand side. Thus we do not have an inequality at this point but an exact equation, but for now we do not specify which diagram the IR leading one might be. The IRE of a diagram can be written as the sum over the IREs of all propagators and all dressed vertices contained, as was done in sec. \ref{ssec:formula-IRE}. The equation can be rewritten using topological relations and transformed into an inequality using the lower bound on IREs given by \eref{eq:Maximal-IR-Divergent}. The general form is:
\ali{\label{eq:leadingTerm}
-\frac1{2}n^b_{{i_1} \ldots {i_r}} \sum_{j} \de_{j} k_{j}^{{i_1} \ldots {i_r}} \geq 0.
}
The derivation of this inequality is presented in Appendix \ref{sec:Prop-Eqs}.
This inequality has decisive differences to the previous ones: The sign is opposite to that in \eref{eq:prop-IRE-inequal1} and it depends on the bare vertex via $n^b_{{i_1} \ldots {i_r}}$. While inequality (\ref{eq:prop-IRE-inequal1}) is always valid, inequality (\ref{eq:leadingTerm}) is only true for the leading diagram, which is not determined up to now. The factor $n^b_{{i_1} \ldots {i_r}}$ divides the large number of possibly leading terms into classes determined by the bare vertex of each diagram as discussed above. For the leading class one can combine eqs. (\ref{eq:prop-IRE-inequal1}) and (\ref{eq:leadingTerm}) to get
\ali{\label{eq:props-equality}
\frac1{2} \hat{n}^b_{{i_1} \ldots {i_r}} \sum_{j}\de_{j} k_{j}^{{i_1} \ldots {i_r}}=0,
}
where one has to keep in mind that the bare vertex determines the appearing IREs. To clarify that this equation is only valid for the leading class we added a hat to $n^b$. This sum is nothing else than adding up the IREs of the propagators corresponding to the legs of a vertex.
From \eref{eq:props-equality} it also follows that the corresponding vertex has a zero IRE, since inequality (\ref{eq:verts-props-inequal1}) reduces to
\ali{
\hat{\de}_{{i_1} \ldots {i_r}}\geq0.
}
Together with the non-positivity of the IREs of primitively divergent vertices, \eref{eq:prim-div}, this yields
\begin{align}
\setlength{\fboxsep}{3mm}
\fbox{$ \displaystyle \hat{\de}_{{i_1} \ldots {i_r}}=0.$}
\end{align}
All one has to do now is to test all possible equalities one can get for a consistent solution. The number of equalities is the same or less than that of the bare vertices of the theory. This is a manageable set of equations that describes the tower of infinitely many coupled DSEs and RGEs.

If the signs of some propagator IREs are known, the number of possible solutions reduces considerably. One can also see that the identity \eref{eq:props-equality}, called scaling relation, can only be fulfilled if a positive and a non-positive IRE are combined, i.~e. having only positive propagator IREs leads to the only solution that all IREs are zero. This corresponds to the perturbative behavior of propagators and vertices in the ultraviolet and seems unlikely to be realized also for small momenta, since perturbation theory itself predicts its breakdown in this regime.

Let us illustrate the efficiency of the method by analyzing the Landau gauge. Here there exist only three different bare vertices, which yield two different equalities:
\ali{
&\de_{gl}=0, & \frac{1}{2}\de_{gl}+\de_{gh}= 0.
}
The first equation leads to the trivial solution $\de_{gl}=\de_{gh}=0$, while the second one yields the known IR scaling solution $\de_g+2\de_{gh}=0$ \cite{vonSmekal:1997vx} with a finite IR dressing function for the ghost-gluon vertex. The IR leading diagrams in the propagator DSEs are those with a bare ghost-gluon vertex and the solution for higher vertex functions \cite{Alkofer:2004it, Huber:2007kc,Fischer:2009tn} can directly be derived by adding to these diagrams further legs using ghost-gluon vertices. This example should make clear that with the method developed here a \textit{complete analysis}, i.~e. including higher vertex functions and two-loop diagrams, can be done in an easy and elegant way without relying on a skeleton expansion or assumptions on the ghost-gluon vertex.

If there are several diagrams with the same bare vertex in the propagator DSEs it may be that more than one solution emerges. The reason is that our analysis assumed that \textit{one} specific diagram is IR leading, but the end result only depends on the bare vertices. Thus one has to distinguish between the results obtained from different diagrams. Note that the obtained scaling relation is not influenced by this, but the behavior of higher vertex functions. Take as an example sunset and squint diagrams. Assuming the squint diagram to be leading, we get that four-point functions are not IR enhanced and thus the sunset is also IR leading. On the other hand starting with the sunset, we do not get information about three-point functions, which are the only dressed vertices in the squint. So it is not clear, whether the squint is IR leading or not in this case. This is exactly what happens in the MAG and consequences are discussed in subsec. \ref{ssec:MAG-SU2}.

One of the main results of this section is given by \eref{eq:props-equality}. In order to get a scaling solution, we write down all realizations of this equation, which are determined by the bare vertices of the theory. Then we test each single equation, if it is consistent with the inequalities derived above. If we find such an equation, it is called a scaling relation for the IREs of the propagators. The vertex corresponding to this equation is called the leading vertex and does not get IR enhanced. The equation may have been derived from several diagrams in the propagator DSEs. If this is the case, details of the solution may depend on it. Indeed the MAG is an example where this happens and details can be found below in subsec. \ref{ssec:MAG-SU2}. The IREs of higher vertex functions are derived by adding further legs to the leading diagram(s) in the propagator DSEs by insertions of the leading vertex.

\section{Infrared scaling solution for the MAG}
\label{sec:IR-Sol-MAG}

Before we employ the method presented in the previous section we will give some details on the MAG in the following subsection. The case of $SU(2)$ is treated then with the method presented above, whereas the inclusion of the additional interactions for higher $SU(N)$ is deferred to an own subsection. Finally we comment on the treatment of the longitudinal part of the propagator for the off-diagonal gluon.

\subsection{The maximally Abelian gauge}

The crucial point of the MAG is the separate treatment of different components of the gauge field. By splitting it up into 
\begin{align}
A_\mu=A_\mu^r T^r=B_\mu^a T^a +A_\mu^i T^i,
\end{align}
where the $T^i$ are the $N-1$ hermitian generators of the $SU(N$) Cartan subalgebra and the $T^a$ the $N^2-N$ remaining generators, we divide it into the so-called diagonal and off-diagonal gluons $A$ and $B$, respectively. This name is based on the possibility to write the generators of the Cartan subalgebra in diagonal form,
\ali{
T^j=\left(\frac{2}{j(j+1)}\right)^{1/2} \times diag(\underbrace{1,\ldots,1}_{j \text{ times}},-j,0,\ldots,0), \quad j=1,\ldots, N-1,
}
whereas the $T^a$ are off-diagonal matrices. Indices are used in accordance with recent literature, i.~e. $a, b, \ldots$ for off-diagonal components and $i,j,\ldots$ for diagonal ones; see e.~g. \cite{Capri:2008ak,Capri:2008vk}. $r,s,\ldots$ refer to both types. We choose Hermitian generators:
\begin{align}\label{eq:commGenerators}
[ T^r,T^s]= i f^{rst}T^t
\end{align}
with the orthogonality condition ($T_f=1/2$ for $SU(N)$)
\begin{align}
tr \lbrace T^r T^s \rbrace = T_f \delta^{rs}.
\end{align}

The underlying idea of the maximally Abelian gauge fixing is to minimize the off-diagonal components \cite{'tHooft:1981ht,Kronfeld:1987ri,Kronfeld:1987vd} and thereby to emphasize the role of the diagonal gluon by demanding that the functional
\begin{align}
R[B]=\int dx\, B_\mu^a B_\mu^a
\end{align}
should be minimized with respect to gauge transformations. This yields the gauge fixing condition
\begin{align}
D_\mu^{ab}B_\mu^b=0
\end{align}
for the off-diagonal gluons with the covariant derivative defined with respect to the diagonal gluon components:
\begin{align}
D_\mu^{ab}:=\delta^{ab} \partial_\mu+g\,f^{abi}A_\mu^i.
\end{align}
For the residual gauge symmetry of the diagonal part we choose Landau gauge, i.~e.
\begin{align}
\partial_\mu A_\mu^i=0.
\end{align}

To keep the gauge renormalizable a quartic ghost interaction term has to be added to the Lagrangian \cite{Min:1985bx,Fazio:2001rm}. All this can be done within the BRST formalism, so that the Lagrangian can be written in the convenient form \cite{Fazio:2001rm}
\begin{align}
\mathcal{L}=&\frac{1}{4}\left( F_{\mu \nu }^aF^a_{\mu \nu}+F_{\mu \nu }^iF^i_{\mu \nu }\right)+s\bar{c}^i(\partial_\mu A_\mu^i-i\frac \xi{2} b^i)+\nnnl
&+s\bar{c}^a(\hat{D}_\mu^{ab} B_\mu^b-i \frac{\alpha}{2} b^a)+s (-\mhalfo \lambda\, g\,f^{abi} \bar{c}^a \bar{c}^b c^i-\frac 1{4}\lambda\, g\,f^{abc} \bar{c}^a \bar{c}^b c^c),
\end{align}
with the BRST transformation $s$ defined as
\begin{align}
s\,A_\mu^a&=-(D_\mu^{ab} c^b-g\,f^{abc}B_\mu^b c^c-g\,f^{abi} B_\mu^b c^i), & s\,A_\mu^i&=-(\delta^{ij}\partial_\mu c_i-g\,f^{iab}B_\mu^a c^b),\nnnl
s\,c^a&=-\mhalfo g\,f^{abc} c^b c^c+g\,f^{abi}c^b c^i, & s\,\bar{c}^a&=i\,b^a, & s\,b^a &=0,\nnnl
s\,c^i&=-\mhalfo g\,f^{iab}c^a c^b, & s\,\bar{c}^i&=i\,b^i, & s\,b^i&=0.
\end{align}
The parameters appearing in the Lagrangian are the gauge fixing parameter for the diagonal part $\xi$ (for Landau gauge it is 0) and that for the off-diagonal part $\alpha$. The parameter $\lambda$ controls the strength of the quartic ghost interaction.
To show explicitly all interactions we give the full form for $SU(N)$, where we introduce a parameter $\zeta$ that interpolates between the MAG ($\zeta=1$) and the Landau gauge ($\zeta=0$):
\begin{align}\label{eq:L}
\mathcal{L}&=\mhalfo(\partial_\mu B_\nu^a)(\partial_\mu B_\nu^a)-\mhalfo(\partial_\mu B_\nu^a)(\partial_\nu B_\mu^a)+\mhalfo g\, f^{abc} B_\mu^b B_\nu^c ((\partial_\mu B_\nu^a)-(\partial_\nu B_\mu^a))+g\, f^{abi}B_\mu^b A_\nu^i ((\partial_\mu B_\nu^a)-(\partial_\nu B_\mu^a))+\nnnl
&+ \mhalfo g^2\, f^{abi}f^{adj}(B_\mu^b A_\nu^i B_\mu^d A_\nu^j-B_\mu^b A_\nu^i A_\mu^j B_\nu^d)+\frac1{4}g^2\,f^{abc} B_\mu^b B_\nu^c (f^{ade} B_\mu^d A_\nu^e +2f^{adi} B_\mu^d A_\nu^i -2 f^{adi} A_\mu^i B_\nu^d)+\nnnl
&+\mhalfo(\partial_\mu A_\nu^i)(\partial_\mu A_\nu^i)-\mhalfo(\partial_\mu A_\nu^i)(\partial_\nu A_\mu^i)+\mhalfo g\,f^{iab}B_\mu^a B_\nu^b((\partial_\mu A_\nu^i)-(\partial_\nu A_\mu^i))+\frac1{4} g^2\,f^{iab} f^{icd}B_\mu^a B_\nu^b B_\mu^c B_\nu^d+\nnnl
&+\bar{c}^a (\delta^{ab} \partial_\mu+\zeta g\,f^{abi}A_\mu^i)(\delta^{bc} \partial_\mu+g\,f^{bcj}A_\mu^j) c^c-g\,f^{abi} \bar{c}^a((\delta^{bc} \partial_\mu+\zeta g\,f^{bcj}A_\mu^j) B_\mu^c) c^i-\nnnl
&-g\,f^{bcd}\bar{c}^a ((\delta^{ab} \partial_\mu+\zeta g\,f^{abi}A_\mu^i) B_\mu^c c^d)-g^2\,\zeta\, f^{abi}f^{cdi} B_\mu^b B_\mu^c \bar{c}^a c^d-
(1-\zeta)g\, f^{abi}\bar{c}^a B_\mu^b \partial_\mu c^i+\nnnl
&+\bar{c}^i\partial_\mu(\partial_\mu c^i-g\,f^{abi}B_\mu^a c^b)+\frac\lambda{4}g^2 f^{abi}f^{cdi}\bar{c}^a\bar{c}^b c^c c^d+\frac{\lambda}{4}g^2 f^{abc}f^{adi}\bar{c}^b \bar{c}^c c^d c^i+\frac{\lambda}{8}g^2 f^{abc}f^{ade} \bar{c}^b \bar{c}^c c^d c^e+\nnnl
&+\frac1{2\alpha}((\delta^{ab} \partial_\mu+\zeta g\,f^{abi}A_\mu^i) B_\mu^b)^2-\frac{\lambda^2}{8\alpha} g^2 f^{abe}f^{cde} \bar{c}^a \bar{c}^c c^b c^d-\frac{\lambda}{\alpha}g\,f^{abi} ((\delta^{ac} \partial_\mu+\zeta g\,f^{acj}A_\mu^j) B_\mu^c) \bar{c}^b c^i-\nnnl
&\frac{\lambda}{2\alpha} g\,f^{abc} ((\delta^{ad} \partial_\mu+\zeta g\,f^{adi}A_\mu^i) B_\mu^d) \bar{c}^b c^c+\frac{\lambda^2}{2\alpha} g^2 f^{aid}f^{bcd} \bar{c}^a \bar{c}^b c^c c^i+\frac1{2\xi}(\partial_\mu A_\mu^i)^2.
\end{align}
The ghost field, being also in the adjoint representation, is split similar to the gluon field. However, the diagonal ghosts decouple for $\lambda=\alpha$, which we will use henceforth. This identity leads to the diagonal ghost equation \cite{Fazio:2001rm}, which is an additional Ward identity respected by the action for $\lambda=\alpha$.

For $SU(2)$ the structure constants assume the simpler form $f^{abc}=0$, $f^{abi}=\epsilon^{ab}$ with $\epsilon$ the antisymmetric tensor, because it is the only possible combination for one diagonal and two off-diagonal fields. For general $SU(N)$ there do not exist non-zero structure constants with more than one diagonal index, as can be seen from \eref{eq:commGenerators}. The Lagrangian contains all types of interactions possible for $SU(N)$. Only in the case of $N=2$ those containing only off-diagonal indices vanish. The interactions are given in table \ref{tab:interactions}.
\begin{table}
\begin{tabular}{l|l|l}
 & Three-Point & Four-Point \\ 
\hline
SU(2) & ABB, Acc & AABB, AAcc, BBBB, cccc, BBcc\\ 
\hline
SU(N) & ABB, Acc Bcc, BBB & AABB, AAcc, BBBB, cccc, BBcc, ABcc, ABBB
\end{tabular}
\caption{\label{tab:interactions}Interactions of the MAG fixed Lagrangian. There are 2 (4) three-point and 5 (7) four-point interactions for $SU(2)$ and $SU(N)$, respectively. $A$ is the diagonal gluon field, $B$ the off-diagonal one and $c$ the ghost field.}
\end{table}

\begin{figure}[th]
\epsfig{file=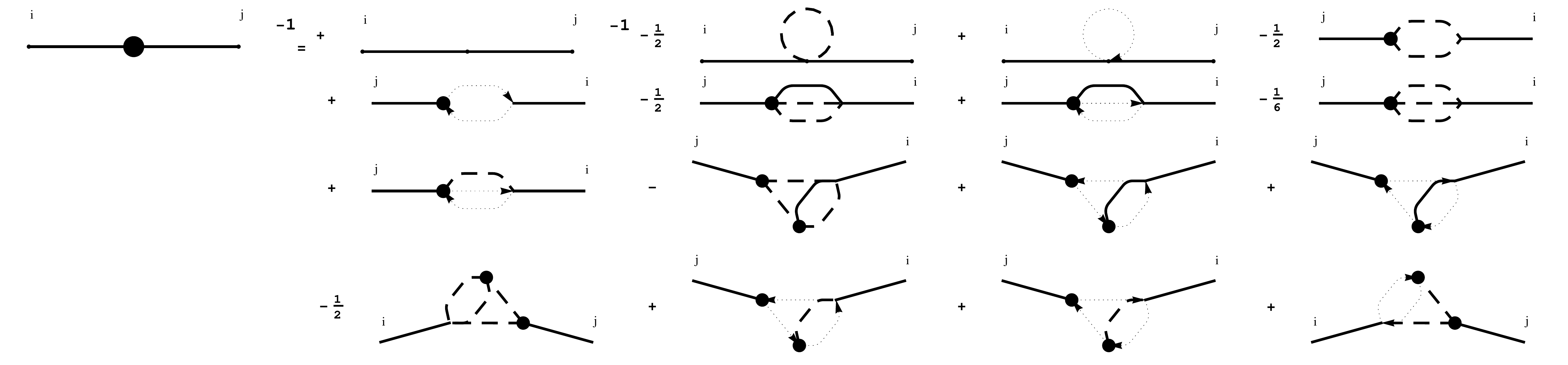,width=\textwidth}
\epsfig{file=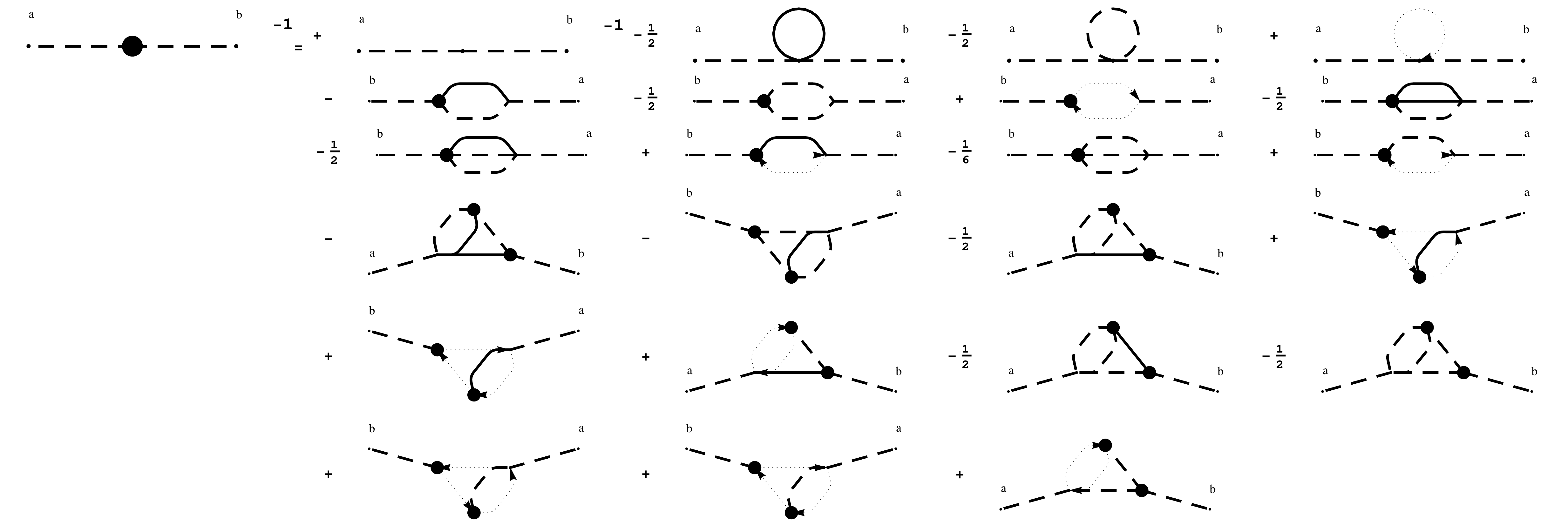,width=\textwidth}
\epsfig{file=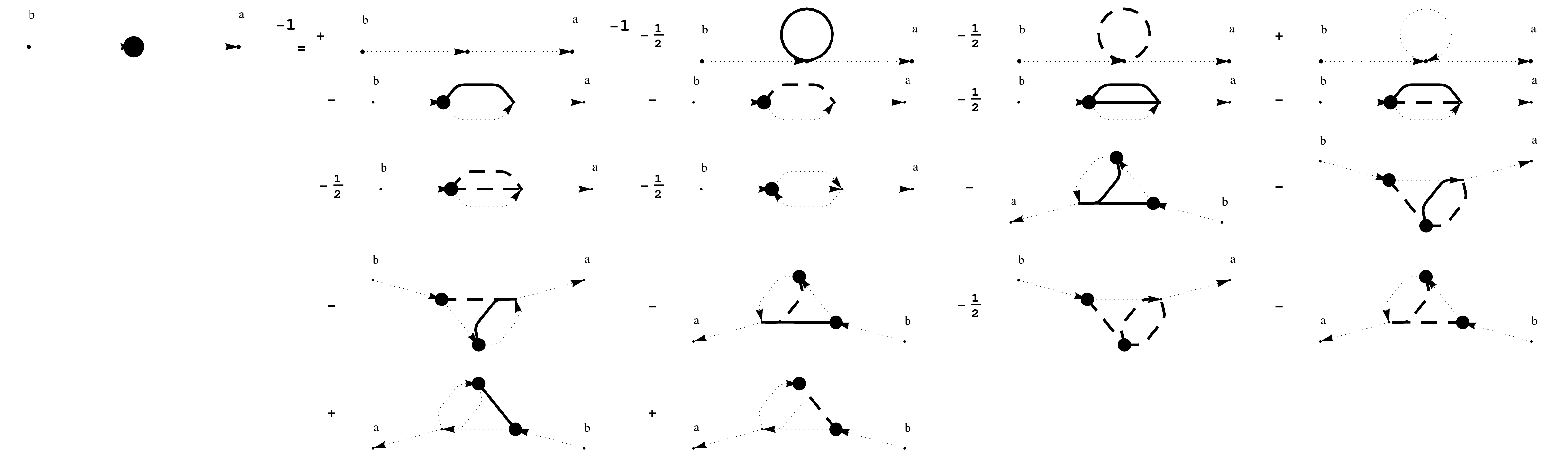,width=\textwidth}
\caption{\label{fig:MAG-DSEs}Yang-Mills DSEs for the propagators in the maximally Abelian gauge. Continuous lines are diagonal gluons, dashed lines off-diagonal gluons and dotted lines ghosts. All internal propagators are dressed, the thick dots are 1PI vertices and the small ones bare vertices. The unusual conventions for the lines are due to the use of \textit{DoDSE} \cite{Alkofer:2008nt}, which cannot draw wavy lines. For the structure functions we did only take into account the antisymmetric $f$-function. Two diagrams (squints with a bare $BBcc$ vertex) that are possible on the combinatoric level do not appear here, because they are zero due to the color algebra.}
\end{figure}

For the derivation of the DSEs we expand the action in the fields, which explicitly shows the types of interactions:
\begin{align}
S=&\mhalfo S^{AA}_{ij} A_i A_j + \mhalfo S^{BB}_{ab} B_a B_b + S^{\bar{c}c}_{ab} \bar{c}_a c_b -
\mhalfo S^{ABB}_{iab} A_i B_a B_b - S^{Acc}_{iab} A_i \bar{c}_a c_b - \frac1{3!} S^{BBB}_{abc} B_a B_b B_c - S^{B\bar{c}c}_{abc} B_a \bar{c}_b c_b+\nnnl
&-\frac1{4} S^{AABB}_{ijab} A_i A_j B_a B_b - \mhalfo S^{AA\bar{c}c}_{ijab} A_i A_j \bar{c}_a c_b - \frac1{4!}S^{BBBB}_{abcd} B_a B_b B_c B_d - \mhalfo S^{BBcc}_{abcd} B_a B_b \bar{c}_c c_d - \frac1{4} S^{\bar{c}\bar{c}cc}_{abcd} \bar{c}_a \bar{c}_b c_c c_d -\nnnl
&- \frac1{3!} S^{ABBB}_{iabc} A_i B_a B_b B_c - S^{ABcc}_{iabc} A_i B_a \bar{c}_b c_c.
\end{align}
The coefficients $S^{\phi_i \ldots \phi_j}_{i\ldots j}$ correspond to the bare n-point functions.
The DSEs were derived with the \textit{Mathematica} package \textit{DoDSE} \cite{Alkofer:2008nt} and are given in \fref{fig:MAG-DSEs}.

Before starting the IR analysis we have to clarify a detail in the DSE of the diagonal gluon propagator: If the bare propagator is not canceled and included in our analysis
its IRE is non-negative. This leads to the trivial solution that all IRE are zero as explained below.
Therefore we argue that for the existence of a scaling solution with IREs unequal to $0$ the propagator DSE of the diagonal gluon may be renormalized in a similar way as the ghost in Landau gauge, so that the bare propagator is absent in the scaling analysis. In Landau gauge this renormalization is connected with the so-called horizon condition, which takes into account the existence of the Gribov horizon \cite{Gribov:1977wm,Zwanziger:1990by,Zwanziger:1991gz,Zwanziger:1993dh,Zwanziger:2001kw,Zwanziger:2002ia,Zwanziger:2003cf}. Recently it was shown in ref. \cite{Capri:2008vk} that the Gribov region in the MAG is quite different from that in the Landau gauge: Whereas in the latter it is bounded in all directions of field space \cite{vanBaal:1991zw,Zwanziger:2003cf}, this is only the case for off-diagonal directions in the MAG \cite{Capri:2008vk}. The diagonal direction on the other hand is unbounded what can lead to a different low energy behavior as in Landau gauge. Another interesting fact is that purely Abelian configurations ($B=0$) that are gauge transformed into Landau gauge lie at the Gribov horizon of Landau gauge \cite{Greensite:2004ke}. But these configurations are exactly the ones important for the IR scaling solution in that gauge.

\subsection{Scaling solution for $SU(2)$}
\label{ssec:MAG-SU2}

With the tools developed above in sec. \ref{sec:General-Approach} the remaining analysis after having established the interactions of the MAG is done rather fast. Introducing the IRE of the diagonal gluon, the off-diagonal gluon and the ghost propagators as $\de_A$, $\de_B$ and $\de_c$, respectively, the occurring interactions give the following list of constraints from the inequalities (\ref{eq:prop-IRE-inequal1}):
\begin{align}\label{eq:inequs-MAG}
\mhalfo \de_A +\de_B \geq & 0, & \de_A +\de_B \geq & 0, & 2\de_B \geq & 0,\nnnl
\mhalfo \de_A +\de_c \geq & 0, & \de_A +\de_c \geq & 0, & 2\de_c \geq & 0,\nnnl
\de_B +\de_c \geq & 0.  &  &  & &
\end{align}
Note that there is a $B\leftrightarrow c$ symmetry in $SU(2)$, i.~e. we could restrict ourselves to only one of these two fields and get the same results. We will not do so in the following.
First we note that both $\de_B$ and $\de_c$ are non-negative\footnote{The corresponding inequalities in $d$ dimensions are $\de_B\geq (4-d)/4$ and $\de_c\geq (4-d)/4$.}, which is a direct consequence of the self-interactions of these fields. This renders the inequality $\de_B +\de_c \geq  0$ superfluous and allows to discard some of the inequalities, since they are less restrictive than others, e.~g. $\frac{1}{2}\de_A+\de_B\geq0$ due to the existence of the $AABB$ interaction that leads to $\de_A+\de_B\geq0$. The remaining system of inequalities for the propagator IREs is
\begin{align}\label{eq:inequs-MAG2}
\de_B \geq & 0, & \de_c \geq & 0, \nnnl
\de_A +\de_B \geq & 0, & \de_A +\de_c \geq & 0.
\end{align}
These inequalities always have to be fulfilled, but at least one of them has to be saturated to have a scaling solution.
We examine every one of the four possibilities.

The first is $\de_B=0$. This renders $\de_A$ also non-negative by virtue of the third inequality. But with only non-negative propagator IREs left, there cannot be a non-trivial scaling solution for the remaining propagators, as this needs positive as well as negative IREs. It is a special feature of the MAG that every field type interacts with every other field type, so that whenever one IRE of a propagator is zero, all others become non-negative and only the trivial solution with $\de_A=\de_B=\de_c=0$ can be realized. Note that there are examples where one field has an IRE of $0$ but the IREs of the other fields still obey a scaling relation, e.~g. massless QCD \cite{Schwenzer:2008vt,Schwenzer:2008fy} or a fundamentally charged massless scalar coupled to Yang-Mills theory in Landau gauge \cite{Fister:2010ah,Fister:2010yw}, but this does not apply for the MAG. In the cited examples this is only possible because the matter fields do not couple to the ghost directly. For $\de_c=0$ the same argument applies, which leaves as only possibility $\de_A +\de_B = 0 $ and $\de_A +\de_c=0$ leading to $\de_{AABB}=\de_{AAcc}=0$.
Here the fields in the subscript determine the legs of the vertex to which the IRE belongs.
It remains to fix the IREs of the other vertices. For vertices with an even number of legs a unique solution exists that corresponds to the maximally IR divergent solution. How this is realized is shown in \fref{fig:higher-vertices}: For a diagram with $n_A$/$n_B$/$n_c$ (all $n_r$ even) legs of type $A$/$B$/$c$ one adds the necessary number of $AABB$ and $AAcc$ vertices to a sunset diagram with a bare $AABB$ or $AAcc$ vertex. Since the added vertices do not scale, the additional internal propagators give exactly the necessary IREs to match the maximally IR divergent solution.
\fig{ht}{fig:higher-vertices}{higher-vertices,width=0.5\textwidth}{Consecutively inserting pairs of $A$, $B$ and $c$ fields shows that for graphs with an even number of legs the maximally IR divergent solution is realized.}

The situation is more intricate for vertices with an odd number of legs, starting already with $ABB$ and $Acc$, which have two possible solutions: The one with the IRE $-\ka/2$, where $\ka:=\de_B=\de_c$, corresponds to the maximally IR divergent solution. The other one has an IRE $0$ for the vertex functions $ABB$ and $Acc$. This duality continues for higher vertex functions, where one can always show that there is an upper bound on the IRE equal to the lower bound plus $\ka/2$. The reason is purely combinatoric: The definitely leading graphs that can be identified unambiguously contain the $AABB$ and $AAcc$ vertices, but it is not possible to construct out of these a diagram with an odd number of legs. Thus, at the end there has to be always at least one vertex with an odd number of legs, which makes the analysis ambiguous.

The IR scaling solution for $SU(2)$ can be summarized as follows:
\begin{itemize}
 \item $-\delta_A=\de_B=\de_c:=\ka \geq0$,
\item $\de(n_A,n_B,n_c)=\mhalfo(n_A-n_B-n_c)\ka \qquad$ ($n_A$ even),
\item $\de(n_A,n_B,n_c)=\mhalfo(n_A-n_B-n_c+\eta)\ka \qquad$ ($n_A$ odd),
\end{itemize}
where $\eta$ can be either $0$ or $1$
and $\de(n_A,n_B,n_c)$ is the IRE of a vertex function with $n_A$ diagonal gluon legs, $n_B$ off-diagonal gluon legs and $n_c$ ghost legs. We introduced the quantity $\ka\geq0$ in analogy to the usual notation in Landau gauge.
For $SU(2)$ the $B$ and $c$ fields only appear in even numbers. Since the IR leading behavior of DSEs is determined by four-point vertices, the dominant contributions arise from two-loop diagrams, namely the sunset and the squint diagrams with a bare $AABB$ or $AAcc$ vertex.
However, there is the possibility that the squint diagrams are IR subleading. This can be traced back to the ambiguity of the vertices with an odd number of legs. Nevertheless the scaling relations $\de_A+\de_B=0$ and $\de_A+\de_c=0$ are the same and determined by the sunset diagrams alone. The question is: Are the squints IR leading as well or IR subleading due to the appearance of two three-point functions?
The latter possibility is an example for a solution where the lower bounds for the IREs are not saturated. It is interesting that in RGEs the leading diagrams in propagator equations are tadpole diagrams in this case.

To understand the ambiguity of vertices with an odd number of legs better we have a closer look at the DSE for the $ABB$ vertex.
It turns out that there are only three types of possibly leading diagrams: Those containing five-point functions (directly coming from the leading sunsets of the propagator equations), those being reduced to $\de_{ABB}$ upon inserting all results known up to now, and those being reduced to $3\de_{ABB}+\ka$. While the former two types are always leading, the latter diagrams only contribute to leading order if the maximally IR divergent solution is realized. On the other hand there are additional diagrams contributing to the leading order if $\de_{ABB}=0$ . The inequalities derived from these diagrams directly correspond to constraints already obtained above (as expected) and do not give any new information. The proof that the diagrams containing five-point functions are definitely leading is done via the DSE of the five-point function. As an example we explain it for the three-point diagram given in \fref{fig:ABB-5-point}. From the three-gluon vertex DSE we have
\ali{
\de_{ABB}\leq \de_{ABBBB}+\ka
}
and from the $ABBBB$ vertex DSE
\ali{
\de_{ABBBB}\leq \de_{ABB} -\ka \quad \Rightarrow \quad \de_{ABB}\geq \de_{ABBBB}+\ka
}
from which we can conclude $\de_{ABB}= \de_{ABBBB}+\ka$.
The duality in the solution for vertices with an odd number of legs does not spoil the equivalence between $\de_{ABB}$ and $\de_{Acc}$, since from the graphs in the corresponding DSEs given in \fref{fig:3-point-ABB-Acc} one has both $\de_{ABB}\leq \de_{Acc}$ and $\de_{Acc}\leq \de_{ABB}$, so that $\de_{ABB}=\de_{Acc}$.

\fig{h}{fig:ABB-5-point}{ABB-5-point,width=0.8\textwidth}{One diagram of the DSE for the three-gluon vertex (left). Part of the DSE for the $ABBBB$ vertex (right).}

\fig{bh}{fig:3-point-ABB-Acc}{3-point-ABB-Acc,width=0.35\textwidth}{Diagrams of the $ABB$ and $Acc$ DSEs that allow to conclude $\de_{ABB}=\de_{Acc}$.}

\subsection{Inclusion of $SU(N)$ interactions}

A special feature of the MAG is its gauge group dependence. It distinguishes it from Landau gauge, where there is no qualitative difference between $SU(2)$ and $SU(3)$ \cite{Sternbeck:2007ug, Cucchieri:2007zm,Maas:2007af}. For the latter exists an additional structure constant in the MAG, which allows additional interaction terms in the Lagrangian. It has three off-diagonal indices, which clearly cannot be provided in $SU(2)$, since there are only two off-diagonal fields. The step from $SU(2)$ to $SU(3)$ may thus change the results from the previous subsection. However, $SU(3)$ already constitutes the most general case, since no other new interactions arise for higher $SU(N)$.

So take into account now the full plethora of interactions of $SU(N)$ Yang-Mills theory. This leads to the following additional constraining inequalities for the propagator IREs:
\begin{align}
\frac3{2}\de_B \geq & 0,  & \mhalfo \de_B +\de_c \geq & 0, &\nnnl
\mhalfo\de_A +\frac3{2}\de_B \geq & 0, & \mhalfo \de_A +\mhalfo \de_B +\de_c \geq & 0. &
\end{align}
They only contain combinations of fields that already occurred, e.~g. $\frac{1}{2}\de_A+\frac{3}{2}\de_B$ can be written as combination of $\frac{1}{2}\de_A+\de_B$ and $\de_B$, of which we already know that they are non-negative. Therefore these inequalities are trivially fulfilled by the solution given above. Furthermore they cannot lead to any new solution, because a scaling solution from any of these terms, e.~g. $1/2\, \de_A + 1/2\, \de_B +\de_c=0$, would directly contradict the $SU(2)$ inequalities. In the given example this would be
\ali{
\de_A+\de_c=(-\de_B-2\de_c)+\de_c=-\de_B-\de_c\geq 0,
}
what directly contradicts the non-negativity of $\de_B$ and $\de_c$, except both are $0$.
Thus for general $SU(N)$ the same diagrams are IR leading as for $SU(2)$.

We established by now that for Green functions with an even number of legs only two-loop diagrams give the IR dominant contributions. This not only demands new methods for a numerical treatment, but also raises the number of dominant diagrams in higher vertex functions considerably, because the vast majority of diagrams possesses a bare four-point function. Already for the ghost-gluon vertex in $SU(2)$ there are only four diagrams out of 27 including the tree-level vertex that are not contributing to the leading order if the maximally IR divergent solution is realized.\footnote{This applies only for that version of the ghost-gluon DSE, where the first differentiation is done with respect to the diagonal gluon.}

We want to remark that the above solution is also valid if we would have started with the Lagrangian without the quartic ghost interaction, so that we would not have $\de_c\geq0$ as a starting point. In this case one can analyze gluons only and establish that the two-loop diagrams with a bare $AABB$ vertex are leading and $\de_A$ is non-positive. Including then the ghosts leads to the result that the $AAcc$ vertex is leading and $\de_c$ is non-negative. Thus the ghost self-interaction is not decisive for the solution but makes the calculation easier, because right from the start it is known that $\de_c\geq0$. Nevertheless this remark is important, because the MAG is only realized by setting the gauge parameter $\alpha$ to $0$ and the bare quartic ghost interaction term is proportional to $\alpha$. Thus some diagrams would vanish from the DSEs\footnote{Note that terms proportional to $1/\alpha$ exist, but they do not appear in the relevant cases to make contributions including a bare quartic ghost vertex.}, and to establish the inequality $\de_c\geq0$ one needs the more involved analysis.

\subsection{Inclusion of the longitudinal dressing function of the $B$ field}

In the analysis above we considered only one dressing function for the off-diagonal gluon propagator. Yet, if the gauge fixing parameter $\alpha$ is not set to zero (which is not possible in general due to terms proportional to $1/\alpha$ in the gluonic vertices), the off-diagonal propagator has a longitudinal part. To take into account its own IRE $\de_{Bl}$ we could add a fourth field and split the vertices accordingly into longitudinal and transverse parts. However, we went even one step further and inserted explicitly the expressions for the tree-level vertices, keeping the dressed vertices general. Projecting longitudinally and transversely, the number of transversal and longitudinal dressing functions is the same for each diagram, i.~e. the equations for the two IREs $\de_{B}$ and $\de_{Bl}$ are equal. This is in contrast to Landau gauge, where the bare three-gluon vertex is zero when contracted with three longitudinal projectors. In the MAG the three-gluon vertex has additional terms that do not vanish. Therefore we conclude that for the analyzed system the two dressing functions have the same IRE.\footnote{This point could be invalidated in case contributions vanish nevertheless when the full (unknown) vertices are inserted into the DSEs. This would imply some yet unknown symmetry of vertex functions not taken into account here.}

\section{Existence of scaling solutions in other gauges}
\label{sec:Other-Gauges}

The method developed in sec. \ref{sec:General-Approach} also allows to investigate other gauges. Again we disregard the possibility of potential cancellations and neglect bare two-point functions in the DSEs. 

\begin{figure}
\includegraphics{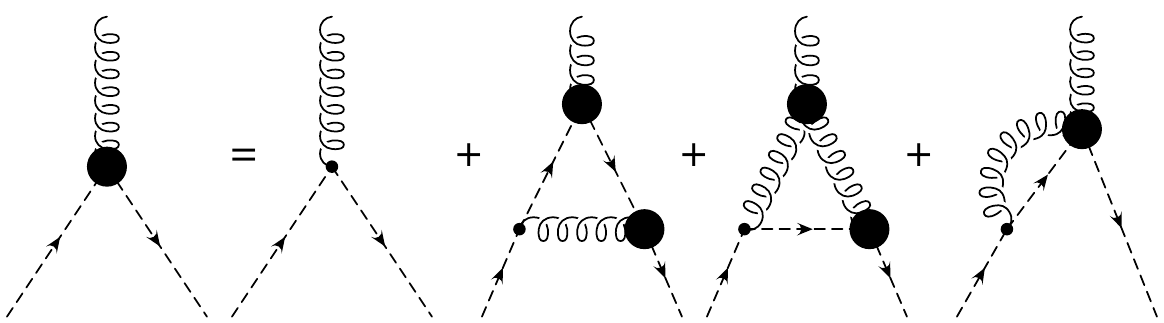}
\caption{\label{fig:ghg-DSE} The ghost-gluon vertex DSE in Landau gauge.}
\end{figure}

The Landau gauge was already used above as an example to show how the method developed in sec. \ref{sec:General-Approach} works and its unique scaling solution \cite{Fischer:2006vf,Fischer:2009tn} was established. We would like to add here an observation concerning the IRE $\de_{gg}$ of the ghost-gluon vertex: Using constraints from \eref{eq:verts-props-inequal1},
\ali{
2\de_{gg}+2\de_{gh}+\de_{gl}\geq0, \quad \de_{gg}+\de_{3g}+\de_{gh}+2\de_{gl}\geq0,
}
in the ghost-gluon vertex DSE where the bare vertices are attached to an external ghost leg, see \fref{fig:ghg-DSE}, one can show that the contributions from the two triangle graphs yield an IRE greater or equal to zero.
A similar constraint derived from \eref{eq:verts-props-inequal1} states that the ghost-gluon scattering kernel cannot be so IR enhanced as to render the ghost-gluon vertex IR divergent:
\ali{
\de_{2gh2g}+\de_{gl}+\de_{gh}\geq0.
}
Thus there is no contribution that is more IR divergent than the bare vertex and the ghost-gluon vertex dressing is IR finite.

A natural extension is the step to general linear covariant gauges away from Landau gauge. In this case the gluon propagator has an additional longitudinal part. Due to gauge invariance this part acquires no dressing and stays bare \cite{Alkofer:2000wg}. However, one can implement the longitudinal part into the presented formalism by a new IRE $\de_{gl,l}$. Since the bare ghost-gluon vertex has a non-vanishing longitudinal part one gets an inequality connecting the longitudinal gluon IRE and the ghost IRE:
\ali{\label{eq:long}
\frac1{2}\de_{gl,l}+\de_{gh}\geq0.
}
The absence of a dressing function for the longitudinal part can be taken into account by setting $\de_{gl,l}$ to zero. Thus one has
\ali{
\de_{gh}\geq0
}
and no scaling relation can be found except the trivial one $\de_{gh}=\de_g=0$. If inequ. (\ref{eq:long}) is indeed valid, the only way to circumvent this is to allow the longitudinal part to get dressed. This leads to $\de_{gl}=\de_{gl,l}=-\frac{1}{2}\de_{gh}$, a result that was already found in ref. \cite{Alkofer:2003jr}.

As a final example we consider ghost anti-ghost symmetric gauges \cite{Baulieu:1981sb,ThierryMieg:1985yv}. In this gauge the longitudinal part of the gluon propagator gets dressed, but there is an additional quartic ghost interaction that again prohibits the existence of a scaling relation, because it leads to the non-negativity of the ghost IRE. As the two gluon IREs are non-negative, too, the only consistent solution is $\de_{gl}=\de_{gl,l}=\de_{gh}=0$. This result was already reported in ref. \cite{Alkofer:2003jr}, where bare four-point vertices were employed for the analysis. Our argument shows that without cancellations a different dressing of the vertices would not cure the failing of a scaling solution.

These two examples allow two possible conclusions: Either the existence of a scaling solution is a special feature of some gauges, or there exist indeed reasons why the constraining inequalities do not hold, e.~g. due to cancellations or the influence of kinematic divergences. In this context it seems promising to pursue interpolating gauges, which allow to investigate how confinement mechanisms in different gauges are linked. Speaking in terms of IREs the interactions that are switched on and off give the relevant information. However, one cannot expect from this point of view that there is a continuous transition, since a non-vanishing interaction always leads to the same scaling relations regardless of the numerical size of its coefficient, which is changed by the  parameters interpolating between different gauges. An example is interpolation between Landau gauge and the MAG, see \eref{eq:L} or ref. \cite{Dudal:2004rx}. The leading vertices in the MAG are the $AABB$ and $AAcc$ vertices. While the two-ghost-two-gluon interaction is absent in Landau gauge at tree-level, the combination of all four-gluon interactions ($AABB$ and $BBBB$) transforms to the standard four-gluon vertex of Landau gauge. Therefore the solution is expected to jump from IR enhanced diagonal gluons to IR enhanced ghosts as one sets the interpolation parameter $\zeta$ in \eref{eq:L} to zero. Whether the picture is really so simple will be investigated in the future.

\section{Conclusions}
\label{sec:Conclusions}

We presented the IR scaling solution for the MAG, which gives a unique qualitative result for the propagators and vertices with an even number of legs. Vertices with an odd number allow several solutions which do not influence the general picture of the IR region, namely that the diagonal gluon is the dominant degree of freedom similar to the ghost propagator in Landau gauge. This behavior is in agreement with the hypothesis of Abelian dominance \cite{Ezawa:1982bf}, which states that the Abelian degrees of freedom should dominate in the IR regime. A necessary condition for the existence of a scaling solution is the cancellation of the bare propagator in the diagonal gluon propagator DSE. We speculated that this might be realized by a similar mechanism as the horizon condition in Landau gauge.

Our analysis is to our knowledge the first investigation of the IR regime of the MAG for the physical gauge group $SU(3)$. However, we did not find a qualitative difference to the simpler case of $SU(2)$, which was not a priori clear due to the higher number of interaction terms in the Lagrangian for $SU(3)$. These terms do not change the scaling solution obtained for $SU(2)$ but only extend it to the new vertices. Furthermore they do not allow for additional solutions. So the qualitative results are obtained for general $SU(N)$ and the emerging structure of DSEs in the IR is the following: The leading diagrams have two loops and a bare vertex that connects two diagonal and two off-diagonal fields. This implies that in general the major part of the diagrams in a DSE contributes to the leading IR order, since with the number of external legs the number of two-loop diagrams grows faster than that of one-loop diagrams.

The method we used is based on the IREs of propagators and vertices and reduces an infinitely large set of DSEs to the relevant information in form of inequalities for the IREs. Thereby we proved that there is a lower bound for the IREs of vertices determined solely by the type and number of legs of the vertices. We called this bound the maximally IR divergent solution. We also showed that a scaling solution always leads to scaling relations linearly connecting the IREs of the propagators, which tells us that there always has to be at least one IR enhanced propagator. However, under certain circumstances (not applying to the MAG) it is possible that only some propagator IREs fulfill a scaling relation and part of the theory remains trivial in the IR.
We want to stress that in the whole analysis we did not truncate the system in any way, but made only two assumptions: The one about the cancellation of the bare diagonal gluon propagator and that there are no cancellations for IR leading terms.

It is also worth mentioning that the method suggested in the present article has been extended successfully to the case of Lagrangians which have a mixing of fields at the level of two-point functions \cite{Huber:2009tx} as it happens, for example, in the Gribov-Zwanziger action \cite{Zwanziger:1989mf}. This action realizes an improved gauge fixing, because the integration in field configuration space is effectively limited to the first Gribov region. It could be shown that this restriction does not interfere with the Dyson-Schwinger analysis \cite{Huber:2009tx}, as already suggested earlier by Zwanziger \cite{Zwanziger:2001kw}, and the obtained solution agrees with the usual result from Faddeev-Popov theory \cite{vonSmekal:1997is,Lerche:2002ep,Zwanziger:2001kw,Pawlowski:2003hq}.

An important point is that our analysis is only concerned with the IR part of the theory. In order to prove that the found solution really exists, it has to be shown in a numerical treatment over the whole momentum regime that it connects to the intermediate and ultraviolet momentum regime.
But it is already a remarkable result that we could find a non-trivial IR scaling solution.
Of course there may be other solutions besides an IR scaling solution as well, like for example the decoupling type of solution found by the refined Gribov-Zwanziger framework \cite{Capri:2008ak} and lattice simulations \cite{Mendes:2008ux}. This issue might be connected with the choice of boundary conditions for the functional equations. In the MAG the boundary condition could be implemented most likely in the DSE of the diagonal gluon propagator by choosing a value for the dressing function at zero momentum. Also the possible presence of additional singularities in different kinematics has not been investigated.

The next step would be of course the numerical solution of the equations. This presents due to the involved dynamics in the MAG a considerable challenge and is therefore left for future studies, which will have as a basis the results obtained here. The fact that two-loop diagrams are important in the IR prohibits the straightforward use of known non-perturbative renormalization methods. It constitutes an unexpected obstacle that is equally or even more complex than the high number of terms in the equations.

We also applied the presented method to other gauges, where it cannot be a priori clear whether a scaling solution exists. Indeed our examples (linear covariant gauges, ghost anti- ghost symmetric gauges) are such, that no scaling solution can be found without further improvements to the employed technique. If thus a scaling solution is specific only to some gauges or if a more detailed analysis would reveal some yet unknown mechanism to allow for a scaling solution also in these gauges has still to be determined. Interpolating gauges might be an interesting application in this context.

\begin{acknowledgments}
We thank David Dudal, Christian S. Fischer, Leonard Fister, Axel Maas, Tereza Mendes, Jan M. Pawlowski, Silvio P. Sorella and Nele Vandersickel for helpful discussions.
K.~S. acknowledges support from the Austrian science fund (FWF) under contract M979-N16 and R.~A. from EU FP7 QCDNet N4. M.~Q.~H. is supported by the Doktoratskolleg ``Hadrons in Vacuum, Nuclei and Stars'' of the FWF under contract W1203-N08.
\end{acknowledgments}

\textbf{Note added in proofs:} As we claimed that this paper is the first to investigate the MAG beyond $SU(2)$, we want to mention that in the meantime an analysis has been performed for $SU(N)$ within the refined Gribov-Zwanziger framework, which confirms the $SU(2)$ results of this approach \cite{Capri:2010an}.

\appendix

\section{Derivation of the IRE formula for an arbitrary diagram}
\label{sec:IRE-Formula}

We derive the formula for the IRE $\delta_v$ of a generic diagram $v$ given in \eref{eq:master-formula}. We start by adding up the exponents of all momenta contributing in an $l$-loop diagram, i.~e. the integral measures give a contribution $l\,d/2$, dressed propagators $(\delta_{{i}}-1)$ and dressed vertices $\delta_{{i_1}\ldots {i_r}}$; canonical dimensions are taken into account by $c_{{i_1}\ldots {i_r}}=2-\frac{1}{2}r$:
\begin{align}\label{eq:deltav-start}
\delta_{v}= & l\frac{d}{2}+\sum_{i}n_i(\delta_{{i}}-1)+\sum_{vertices,r\geq3}n^{d}_{{i_1}\ldots {i_r}}(\delta_{{i_1}\ldots {i_r}}+c_{{i_1}\ldots {i_r}})+\nnnl
 & +\sum_{vertices,r\geq3}n^{b}_{{i_1}\ldots {i_r}}c_{{i_1}\ldots {i_r}}-c_{v}.
\end{align}
$n_i$ are the number of internal propagators with IREs $\delta_{{i}}$, whereas the numbers of vertices are $n^{{i_1}\ldots {i_r}}$. Superscripts $d$ and $b$ stand for dressed and bare, respectively. In case none is given, we refer to both. The sums $\sum_{vertices,r\geq3}$ extend over all vertices with at most $r$ legs; $m_{i}$ being the number of external legs, the maximal value is $\sum_{i}m_{i}+2$, since the highest vertex functions in DSEs and RGEs have at most two legs more than the vertex function on the left-hand side of the equation.

We will use the following topological relations to rewrite the equation above: The number of vertices and propagators of a certain field $\phi_i$ are connected by
\begin{align}\label{eq:relation-verts-props}
n_i=\frac{1}{2}\left(\sum_{vertices,r\geq3}k_{{i}}^{{i_1}\ldots {i_r}}n_{{i_1}\ldots {i_r}}-m_{i}\right),
\end{align}
where $k_{{i}}^{{i_1}\ldots {i_r}}$ denotes the number of times the field $\phi_{i}$ appears in the vertex $\phi_{i_1}\ldots \phi_{i_r}$. The number of loops can be replaced using
\begin{align}
l=\sum_{i}n_i+1-\sum_{vertices,r\geq3}n_{{i_1}\ldots {i_r}}.
\end{align}
Plugging these expressions into \eref{eq:deltav-start} yields
\begin{align}
\delta_{v} = & \left(\sum_{i}\frac{1}{2}\left(\sum_{vertices,r\geq3}k_{{i}}^{{i_1}\ldots {i_r}}n_{{i_1}\ldots {i_r}}-m_{i}\right)+1-\sum_{vertices,r\geq3}n_{{i_1}\ldots {i_r}}\right)\frac{d}{2}+\nnnl
 & +\sum_{i}\frac{1}{2}\left(\sum_{vertices,r\geq3}k_{{i}}^{{i_1}\ldots {i_r}}n_{{i_1}\ldots {i_r}}-m_{i}\right)(\delta_{{i}}-1)+\nnnl
 & +\sum_{vertices,r\geq3}n^{d}_{{i_1}\ldots {i_r}}\left(\delta_{{i_1}\ldots {i_r}}+2-\frac{r}{2}\right)+\sum_{vertices,r\geq3}n^{b}_{{i_1}\ldots {i_r}}\left(2-\frac{r}{2}\right)-2+\frac{1}{2}\sum_{i}m_{i}=\nnnl
= & \left(\frac{d}{2}-2\right)\left(1-\frac{1}{2}\sum_{i}m_{i}\right)-\frac{1}{2}\sum_{i}m_{i}\delta_{{i}}+
 \sum_{vertices,r\geq3}n^{d}_{{i_1}\ldots {i_r}}\left(-\frac{d}{2}+\delta_{{i_1}\ldots {i_r}}+2-\frac{r}{2}\right)+\nnnl
 & +\sum_{i}\frac{1}{2}\left(\sum_{vertices,r\geq3}k_{{i}}^{{i_1}\ldots {i_r}}n_{{i_1}\ldots {i_r}}(\frac{d}{2}+\delta_{{i}}-1)\right)+
 \sum_{vertices,r\geq3}n^{b}_{{i_1}\ldots {i_r}}\left(-\frac{d}{2}+2-\frac{r}{2}\right).
\end{align}
Using $\sum_{i}k_{{i}}^{{i_1}\ldots {i_r}}=r$ leads to
\begin{align}\label{eq:master-formula-d}
\delta_{v} = & \left(\frac{d}{2}-2\right)\left(1-\frac{1}{2}\sum_{i}m_{i}\right)-\frac{1}{2}\sum_{i}m_{i}\delta_{{i}}+\nnnl
 & +\sum_{vertices,r\geq3}n^{d}_{{i_1}\ldots {i_r}}\left(\left(\frac{d}{4}-1\right)(r-2)+\delta_{{i_1}\ldots {i_r}}+\frac{1}{2}\sum_{i}k_{{i}}^{{i_1}\ldots {i_r}}\delta_{{i}}\right)+\nnnl
 & +\sum_{vertices,r\geq3}n^{b}_{{i_1}\ldots {i_r}}\left(\left(\frac{d}{4}-1\right)(r-2)+\frac{1}{2}\sum_{i}k_{{i}}^{{i_1}\ldots {i_r}}\delta_{{i}}\right).
\end{align}
In four dimensions this result reduces to \eref{eq:master-formula}.

\section{Constraints from RGEs}
\label{sec:Constraints}

We outline the derivation of the constraints for four- and higher n-point functions and include the dimension $d$ explicitly. The resulting expressions match exactly the coefficients of $n^d_{{i_1} \ldots {i_r}}$ in \eref{eq:master-formula}.

\begin{figure}[b]
\begin{center}
 a) \includegraphics[width=0.43\textwidth]{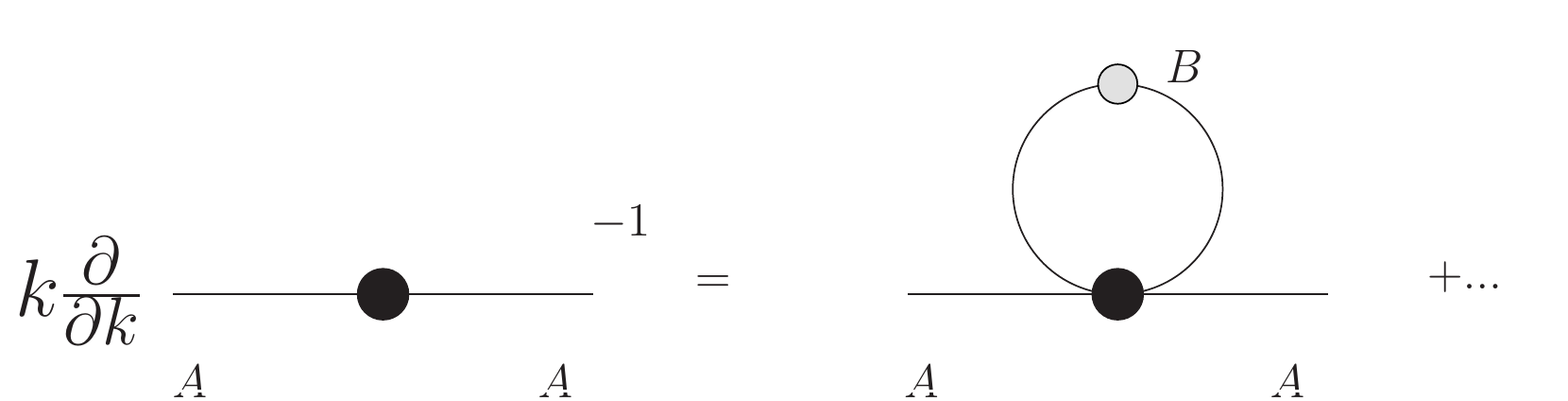}\hskip1cm
 b) \includegraphics[width=0.43\textwidth]{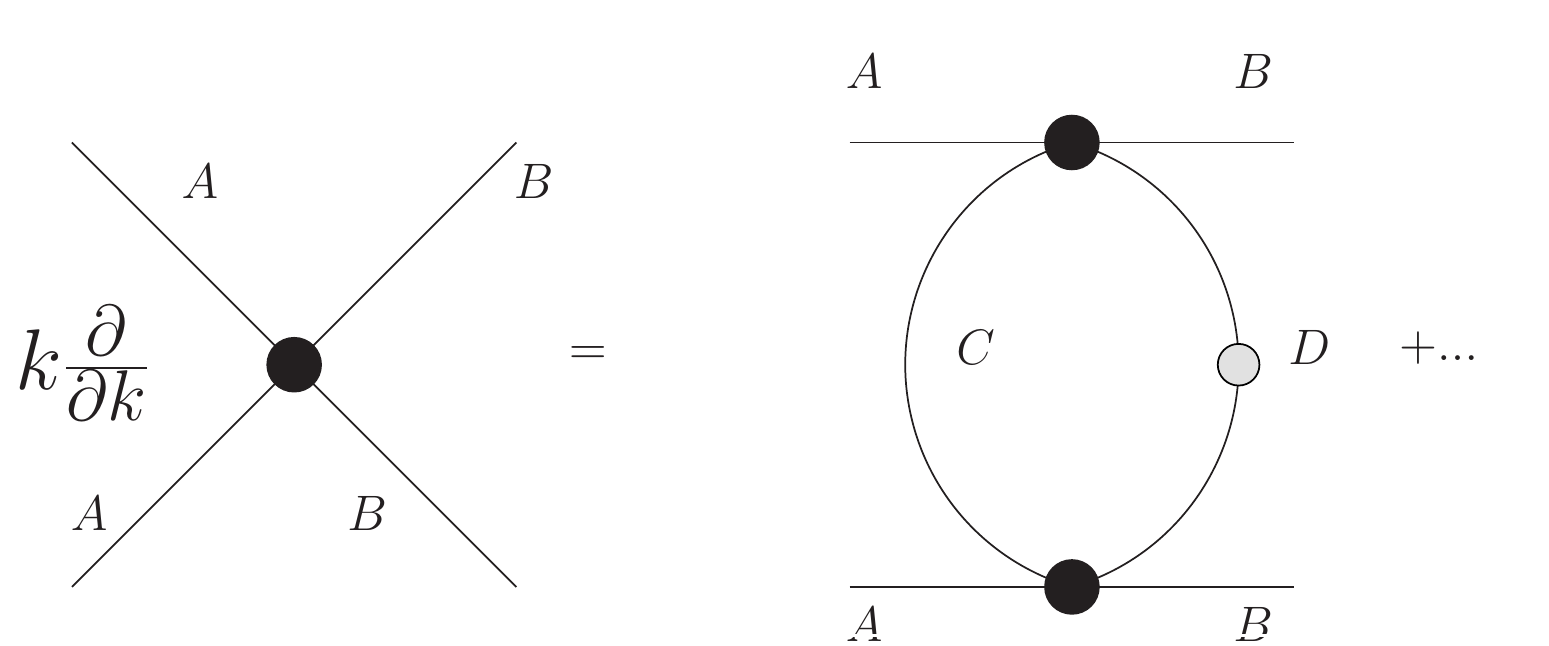}
\caption{\label{fig:RG-4-point-App} Parts of the FRGEs of generic four-point functions.}
\end{center}
\end{figure}

For four-point functions we get from \fref{fig:RG-4-point-App} the following two inequalities:
\begin{align}
 -\delta_A-\delta_B -\left(\frac{d}{2}-2\right) &\leq \de_{AABB},\label{eq:4Point1}\\
 \de_{AABB}&\leq 2\de_{ABCD}+\delta_{C}+\delta_{D}+\frac{d}{2}-2 \label{eq:4Point2}.
\end{align}
Combining them yields
\begin{align}
 \de_{ABCD}+\frac{1}{2} \left(\delta_{A}+\delta_{B}+\delta_{C}+\delta_{D}\right) +2\left(\frac{d}{4}-1\right)\geq0.
\end{align}

The inequalities for the three- and four-point functions fulfill
\begin{align}\label{eq:IR-ineqs1-App}
\left(\frac{d}{4}-1\right)(r-2)+\de_{{i_1}\ldots {i_r}}+\frac{1}{2}\sum_{j}k_{{j}}^{{i_1}\ldots {i_r}}\delta_{{i}}\geq0,
\end{align}
where $r=3,4$ denotes the number of legs the n-point function has.
Higher n-point function also obey this inequality as can be shown by induction.

\begin{figure}[t]
\begin{center}
 a) \includegraphics[width=0.43\textwidth]{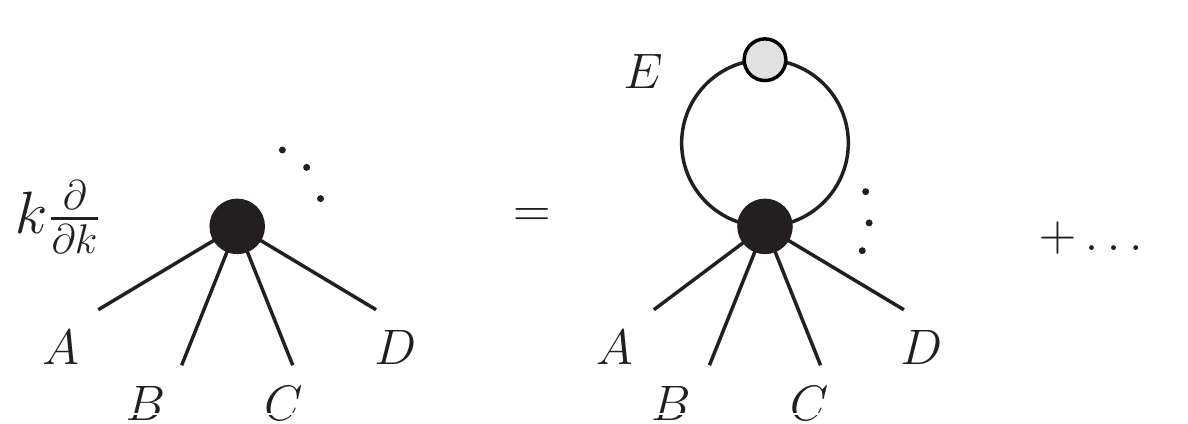}\hskip1cm
 b) \includegraphics[width=0.43\textwidth]{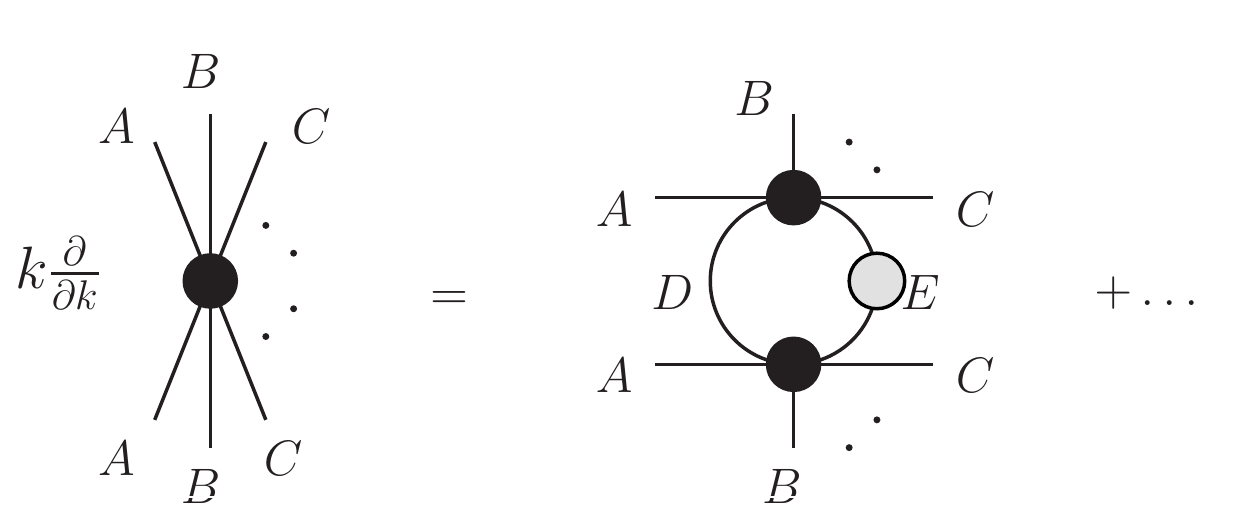}
\caption{\label{fig:RG-3+6+-app} Parts of the FRGEs of generic n-point functions.}
\end{center}
\end{figure}

For the proof we need the following two inequalities which can be inferred from \fref{fig:RG-3+6+-app}:
\begin{align}
 \de_{ABCD\cdots}&\leq \de_{ABCDEE\cdots}+\de_E+\frac{d}{2}-2,\label{eq:nPoint1-App}\\
 \de_{AABBCC\cdots}&\leq 2\de_{ABCDE\cdots}+\de_D+\de_E+\frac{d}{2}-2.
\end{align}
The dots represent further legs as indicated in  \fref{fig:RG-3+6+-app}. Note that these inequalities are generalizations of eqs. (\ref{eq:4Point1}) and (\ref{eq:4Point2}).
The first inequality can be used to write down an equation for the vertex $AABBCC\ldots$ as appearing in the second inequality:
\begin{align}
 \de_{AABB\cdots}-\de_{C}-\left(\frac{d}{2}-2\right)\leq\de_{AABBCC\cdots}&\leq 2\de_{ABCDE\cdots}+\de_D+\de_E+\frac{d}{2}-2,\\
 \frac{1}{2}\left(\de_{AABB\cdots}-\de_C-\de_D-\de_E\right)-\left(\frac{d}{2}-2\right) &\leq \de_{ABCDE\cdots}\,\label{eq:nPoint2-App}.
\end{align}
This inequality connects an n-point function with a (2n-6)-point function. The goal is to rewrite the equation such that only propagator IREs and the IRE of the n-point function remain. For this one can successively use \eref{eq:nPoint1-App} to replace the IRE of the remaining other vertex and arrives at
\begin{align}
 -\frac{1}{2}\left(\de_A+\de_B+\de_C+\de_D+\de_E +\dots \right)-\left(\frac{d}{4}-1\right)(r-2) &\leq \de_{ABCDE\cdots}\,,
\end{align}
where $r$ gives the number of legs of the n-point function. The coefficient of $d/4-1$ is obtained as follows: One gets $2(d/4-1)$ in \eref{eq:nPoint2-App} and $d/4-1$ from every further application of \eref{eq:nPoint1-App}. The latter has to be done until $2n-6-2x=2$, where $x$ is the number of iterations. Hence the total coefficient of $d/4-1$ is $2+x=n-2$.  This establishes \eref{eq:IR-ineqs1-App} for all n-point functions.

\section{Inequality from the leading diagram}
\label{sec:Prop-Eqs}

We give here the derivation of \eref{eq:leadingTerm}.
For the leading diagram in a propagator DSE we get from \eref{eq:master-formula-d}
\ali{\label{eq:leading-diagram}
-\de_{i}&=\left(\frac{d}{2}-2\right)\left(1-\frac1{2}\sum_i m_i \right)-\frac1{2}\sum_i m_i \de_i+\nnnl
&+\sum_{\substack{vertices}} n_{{i_1} \ldots {i_r}} \left(\left(\frac{d}{4}-1\right)(r-2)+\frac1{2}\sum_i k_i^{i_1 \ldots i_r} \de_i \right)
+\sum_{\substack{dressed\\vertices}}  n^d_{{i_1} \ldots {i_r}} \de_{{i_1} \ldots {i_r}}.
}
Using $\sum_i m_i=2$ and $\sum_i m_i \de_i=2\de_i$ this can be written as
\begin{align}
-\sum_{\substack{vertices}} n_{{i_1} \ldots {i_r}} \left(\left(\frac{d}{4}-1\right)(r-2)+\frac1{2}\sum_i k_i^{i_1 \ldots i_r} \de_i \right)=
\sum_{\substack{dressed\\vertices}}  n^d_{{i_1} \ldots {i_r}} \de_{{i_1} \ldots {i_r}}.
\end{align}
This is a general relation for the leading diagram, whichever this might be.
Now we express the IREs of the vertices, $\de_{{i_1} \ldots {i_r}}$, by their lower bound, derived from \eref{eq:master-formula-d}, to arrive at a new inequality:
\begin{align}
-\sum_{\substack{vertices}}& n_{{i_1} \ldots {i_r}} \left(\left(\frac{d}{4}-1\right)(r-2)+\frac1{2}\sum_i k_i^{i_1 \ldots i_r} \de_i \right)\geq \nnnl
 &\geq\sum_{\substack{dressed\\vertices}} n^d_{{i_1} \ldots {i_r}} \left( -\frac1{2}\sum_j \de_{j} k_{j}^{{i_1} \ldots {i_r}} +\left(\frac{d}{2}-2\right)\left(1-\frac1{2}\sum_i k_{i}^{{i_1} \ldots {i_r}} \right) \right).
\end{align}
The right-hand side depends on dressed vertices only, indicated by the $d$ superscript of $n$. On the other hand the left-hand side sums over dressed and bare vertices, so that in total only the bare vertex remains:
\begin{align}
- n^b_{{i_1} \ldots {i_r}} \left(\left(\frac{d}{4}-1\right)(r-2)+\frac1{2}\sum_i k_i^{i_1 \ldots i_r} \de_i \right)\geq0,
\end{align}
where we used $\sum_i k_i^{i_1 \ldots i_r}=r$.
For $d=4$ this is exactly \eref{eq:leadingTerm}.

\bibliographystyle{utphys}
\bibliography{literature}

\providecommand{\href}[2]{#2}\begingroup\raggedright\begin{thebibliography}{10}

\bibitem{Alkofer:2000wg}
R.~Alkofer and L.~von Smekal, {\em Phys. Rept.} {\bf 353} (2001)  281,
\href{http://arxiv.org/abs/hep-ph/0007355}{{\tt hep-ph/0007355}}.

\bibitem{Fischer:2006ub}
C.~S. Fischer, {\em J. Phys.} {\bf G32} (2006)  R253--R291,
\href{http://arxiv.org/abs/hep-ph/0605173}{{\tt hep-ph/0605173}}.

\bibitem{Berges:2000ew}
J.~Berges, N.~Tetradis, and C.~Wetterich, {\em Phys. Rept.} {\bf 363} (2002)
  223--386,
\href{http://arxiv.org/abs/hep-ph/0005122}{{\tt arXiv:hep-ph/0005122}}.

\bibitem{Pawlowski:2005xe}
J.~M. Pawlowski, \href{http://dx.doi.org/10.1016/j.aop.2007.01.007}{{\em Annals
  Phys.} {\bf 322} (2007)  2831--2915},
\href{http://arxiv.org/abs/hep-th/0512261}{{\tt arXiv:hep-th/0512261}}.

\bibitem{vonSmekal:1997vx}
L.~von Smekal, A.~Hauck, and R.~Alkofer,
  \href{http://dx.doi.org/10.1006/aphy.1998.5806}{{\em Ann. Phys.} {\bf 267}
  (1998)  1},
\href{http://arxiv.org/abs/hep-ph/9707327}{{\tt arXiv:hep-ph/9707327}}.

\bibitem{vonSmekal:1997is}
L.~von Smekal, R.~Alkofer, and A.~Hauck,
  \href{http://dx.doi.org/10.1103/PhysRevLett.79.3591}{{\em Phys. Rev. Lett.}
  {\bf 79} (1997)  3591--3594},
\href{http://arxiv.org/abs/hep-ph/9705242}{{\tt arXiv:hep-ph/9705242}}.

\bibitem{Pawlowski:2003hq}
J.~M. Pawlowski, D.~F. Litim, S.~Nedelko, and L.~von Smekal, {\em Phys. Rev.
  Lett.} {\bf 93} (2004)  152002,
\href{http://arxiv.org/abs/hep-th/0312324}{{\tt hep-th/0312324}}.

\bibitem{Zwanziger:2001kw}
D.~Zwanziger, {\em Phys. Rev.} {\bf D65} (2002)  094039,
\href{http://arxiv.org/abs/hep-th/0109224}{{\tt hep-th/0109224}}.

\bibitem{Zwanziger:2002ia}
D.~Zwanziger, {\em Phys. Rev.} {\bf D67} (2003)  105001,
\href{http://arxiv.org/abs/hep-th/0206053}{{\tt hep-th/0206053}}.

\bibitem{Zwanziger:2003cf}
D.~Zwanziger, \href{http://dx.doi.org/10.1103/PhysRevD.69.016002}{{\em Phys.
  Rev.} {\bf D69} (2004)  016002},
\href{http://arxiv.org/abs/hep-ph/0303028}{{\tt arXiv:hep-ph/0303028}}.

\bibitem{Lerche:2002ep}
C.~Lerche and L.~von Smekal, {\em Phys. Rev.} {\bf D65} (2002)  125006,
\href{http://arxiv.org/abs/hep-ph/0202194}{{\tt hep-ph/0202194}}.

\bibitem{Huber:2009tx}
M.~Q. Huber, R.~Alkofer, and S.~P. Sorella,
  \href{http://dx.doi.org/10.1103/PhysRevD.81.065003}{{\em Phys. Rev.} {\bf
  D81} (2010)  065003},
\href{http://arxiv.org/abs/0910.5604}{{\tt arXiv:0910.5604 [hep-th]}}.

\bibitem{Alkofer:2006gz}
R.~Alkofer, C.~S. Fischer, and F.~J. Llanes-Estrada,
  \href{http://dx.doi.org/10.1142/S021773230802700X}{{\em Mod. Phys. Lett.}
  {\bf A23} (2008)  1105},
\href{http://arxiv.org/abs/hep-ph/0607293}{{\tt arXiv:hep-ph/0607293}}.

\bibitem{Alkofer:2008tt}
R.~Alkofer, C.~S. Fischer, F.~J. Llanes-Estrada, and K.~Schwenzer,
  \href{http://dx.doi.org/10.1016/j.aop.2008.07.001}{{\em Annals Phys.} {\bf
  324} (2009)  106--172},
\href{http://arxiv.org/abs/0804.3042}{{\tt arXiv:0804.3042 [hep-ph]}}.

\bibitem{Taylor:1971ff}
J.~C. Taylor,
{\em Nucl. Phys.} {\bf B33} (1971)  436--444.

\bibitem{Schleifenbaum:2004id}
W.~Schleifenbaum, A.~Maas, J.~Wambach, and R.~Alkofer, {\em Phys. Rev.} {\bf
  D72} (2005)  014017,
\href{http://arxiv.org/abs/hep-ph/0411052}{{\tt hep-ph/0411052}}.

\bibitem{Alkofer:2008jy}
R.~Alkofer, M.~Q. Huber, and K.~Schwenzer,
  \href{http://dx.doi.org/http://link.aps.org/doi/10.1103/PhysRevD.81.105010}{%
{\em Phys. Rev.} {\bf D81} (2010)  105010},
\href{http://arxiv.org/abs/0801.2762}{{\tt arXiv:0801.2762 [hep-th]}}.

\bibitem{Fischer:2006vf}
C.~S. Fischer and J.~M. Pawlowski,
  \href{http://dx.doi.org/10.1103/PhysRevD.75.025012}{{\em Phys. Rev.} {\bf
  D75} (2007)  025012},
\href{http://arxiv.org/abs/hep-th/0609009}{{\tt arXiv:hep-th/0609009}}.

\bibitem{Fischer:2009tn}
C.~S. Fischer and J.~M. Pawlowski,
  \href{http://dx.doi.org/10.1103/PhysRevD.80.025023}{{\em Phys. Rev.} {\bf
  D80} (2009)  025023},
\href{http://arxiv.org/abs/0903.2193}{{\tt arXiv:0903.2193 [hep-th]}}.

\bibitem{Cucchieri:2004sq}
A.~Cucchieri, T.~Mendes, and A.~Mihara, {\em JHEP} {\bf 12} (2004)  012,
\href{http://arxiv.org/abs/hep-lat/0408034}{{\tt hep-lat/0408034}}.

\bibitem{Ilgenfritz:2006he}
E.~M. Ilgenfritz, M.~Muller-Preussker, A.~Sternbeck, A.~Schiller, and I.~L.
  Bogolubsky, {\em Braz. J. Phys.} {\bf 37} (2007)  193,
\href{http://arxiv.org/abs/hep-lat/0609043}{{\tt arXiv:hep-lat/0609043}}.

\bibitem{Cucchieri:2008qm}
A.~Cucchieri, A.~Maas, and T.~Mendes,
  \href{http://dx.doi.org/10.1103/PhysRevD.77.094510}{{\em Phys. Rev.} {\bf
  D77} (2008)  094510},
\href{http://arxiv.org/abs/0803.1798}{{\tt arXiv:0803.1798 [hep-lat]}}.

\bibitem{Fischer:2003rp}
C.~S. Fischer and R.~Alkofer, {\em Phys. Rev.} {\bf D67} (2003)  094020,
\href{http://arxiv.org/abs/hep-ph/0301094}{{\tt hep-ph/0301094}}.

\bibitem{Fischer:2008uz}
C.~S. Fischer, A.~Maas, and J.~M. Pawlowski,
  \href{http://dx.doi.org/10.1016/j.aop.2009.07.009}{{\em Annals Phys.} {\bf
  324} (2009)  2408--2437},
\href{http://arxiv.org/abs/0810.1987}{{\tt arXiv:0810.1987 [hep-ph]}}.

\bibitem{Alkofer:2008dt}
R.~Alkofer, M.~Q. Huber, and K.~Schwenzer,
  \href{http://dx.doi.org/10.1140/epjc/s10052-009-1066-3}{{\em Eur. Phys. J.}
  {\bf C62} (2009)  761--781},
\href{http://arxiv.org/abs/0812.4045}{{\tt arXiv:0812.4045 [hep-ph]}}.

\bibitem{Kellermann:2008iw}
C.~Kellermann and C.~S. Fischer,
  \href{http://dx.doi.org/10.1103/PhysRevD.78.025015}{{\em Phys. Rev.} {\bf
  D78} (2008)  025015},
\href{http://arxiv.org/abs/0801.2697}{{\tt arXiv:0801.2697 [hep-ph]}}.

\bibitem{Alkofer:2004it}
R.~Alkofer, C.~S. Fischer, and F.~J. Llanes-Estrada, {\em Phys. Lett.} {\bf
  B611} (2005)  279--288,
\href{http://arxiv.org/abs/hep-th/0412330}{{\tt hep-th/0412330}}.

\bibitem{Huber:2007kc}
M.~Q. Huber, R.~Alkofer, C.~S. Fischer, and K.~Schwenzer,
  \href{http://dx.doi.org/10.1016/j.physletb.2007.10.073}{{\em Phys. Lett.}
  {\bf B659} (2008)  434--440},
\href{http://arxiv.org/abs/0705.3809}{{\tt arXiv:0705.3809 [hep-ph]}}.

\bibitem{Alkofer:2007hc}
R.~Alkofer, C.~S. Fischer, M.~Q. Huber, and K.~Schwenzer, {\em PoS} {\bf
  LAT2007} (2007)  329,
\href{http://arxiv.org/abs/0710.1054}{{\tt arXiv:0710.1054 [hep-ph]}}.

\bibitem{Boucaud:2008ji}
P.~Boucaud {\em et al.},
  \href{http://dx.doi.org/10.1088/1126-6708/2008/06/012}{{\em JHEP} {\bf 06}
  (2008)  012},
\href{http://arxiv.org/abs/0801.2721}{{\tt arXiv:0801.2721 [hep-ph]}}.

\bibitem{Aguilar:2008xm}
A.~C. Aguilar, D.~Binosi, and J.~Papavassiliou,
  \href{http://dx.doi.org/10.1103/PhysRevD.78.025010}{{\em Phys. Rev.} {\bf
  D78} (2008)  025010},
\href{http://arxiv.org/abs/0802.1870}{{\tt arXiv:0802.1870 [hep-ph]}}.

\bibitem{Dudal:2007cw}
D.~Dudal, S.~P. Sorella, N.~Vandersickel, and H.~Verschelde,
  \href{http://dx.doi.org/10.1103/PhysRevD.77.071501}{{\em Phys. Rev.} {\bf
  D77} (2008)  071501},
\href{http://arxiv.org/abs/0711.4496}{{\tt arXiv:0711.4496 [hep-th]}}.

\bibitem{Braun:2007bx}
J.~Braun, H.~Gies, and J.~M. Pawlowski,
  \href{http://dx.doi.org/10.1016/j.physletb.2010.01.009}{{\em Phys. Lett.}
  {\bf B684} (2010)  262--267},
\href{http://arxiv.org/abs/0708.2413}{{\tt arXiv:0708.2413 [hep-th]}}.

\bibitem{Cucchieri:2007md}
A.~Cucchieri and T.~Mendes, {\em PoS} {\bf LAT2007} (2007)  297,
\href{http://arxiv.org/abs/0710.0412}{{\tt arXiv:0710.0412 [hep-lat]}}.

\bibitem{Cucchieri:2007rg}
A.~Cucchieri and T.~Mendes,
  \href{http://dx.doi.org/10.1103/PhysRevLett.100.241601}{{\em Phys. Rev.
  Lett.} {\bf 100} (2008)  241601},
\href{http://arxiv.org/abs/0712.3517}{{\tt arXiv:0712.3517 [hep-lat]}}.

\bibitem{Cucchieri:2008fc}
A.~Cucchieri and T.~Mendes,
  \href{http://dx.doi.org/10.1103/PhysRevD.78.094503}{{\em Phys. Rev.} {\bf
  D78} (2008)  094503},
\href{http://arxiv.org/abs/0804.2371}{{\tt arXiv:0804.2371 [hep-lat]}}.

\bibitem{Bogolubsky:2009dc}
I.~L. Bogolubsky, E.~M. Ilgenfritz, M.~Muller-Preussker, and A.~Sternbeck,
  \href{http://dx.doi.org/10.1016/j.physletb.2009.04.076}{{\em Phys. Lett.}
  {\bf B676} (2009)  69--73},
\href{http://arxiv.org/abs/0901.0736}{{\tt arXiv:0901.0736 [hep-lat]}}.

\bibitem{Bornyakov:2008yx}
V.~G. Bornyakov, V.~K. Mitrjushkin, and M.~Muller-Preussker,
  \href{http://dx.doi.org/10.1103/PhysRevD.79.074504}{{\em Phys. Rev.} {\bf
  D79} (2009)  074504},
\href{http://arxiv.org/abs/0812.2761}{{\tt arXiv:0812.2761 [hep-lat]}}.

\bibitem{Pawlowski:2009iv}
J.~M. Pawlowski, D.~Spielmann, and I.-O. Stamatescu,
  \href{http://dx.doi.org/10.1016/j.nuclphysb.2009.12.036}{{\em Nucl. Phys.}
  {\bf B830} (2010)  291--314},
\href{http://arxiv.org/abs/0911.4921}{{\tt arXiv:0911.4921 [hep-lat]}}.

\bibitem{Dudal:2008sp}
D.~Dudal, J.~A. Gracey, S.~P. Sorella, N.~Vandersickel, and H.~Verschelde,
  \href{http://dx.doi.org/10.1103/PhysRevD.78.065047}{{\em Phys. Rev.} {\bf
  D78} (2008)  065047},
\href{http://arxiv.org/abs/0806.4348}{{\tt arXiv:0806.4348 [hep-th]}}.

\bibitem{Sternbeck:2008na}
A.~Sternbeck and L.~von Smekal, {\em PoS} {\bf Confinement8} (2008)  049,
\href{http://arxiv.org/abs/0812.3268}{{\tt arXiv:0812.3268 [hep-lat]}}.

\bibitem{Maas:2008ri}
A.~Maas, \href{http://dx.doi.org/10.1103/PhysRevD.79.014505}{{\em Phys. Rev.}
  {\bf D79} (2009)  014505},
\href{http://arxiv.org/abs/0808.3047}{{\tt arXiv:0808.3047 [hep-lat]}}.

\bibitem{Maas:2009se}
A.~Maas, \href{http://dx.doi.org/10.1016/j.physletb.2010.04.052}{{\em Phys.
  Lett.} {\bf B689} (2010)  107--111},
\href{http://arxiv.org/abs/0907.5185}{{\tt arXiv:0907.5185 [hep-lat]}}.

\bibitem{Mandelstam:1974pi}
S.~Mandelstam,
\href{http://dx.doi.org/10.1016/0370-1573(76)90043-0}{{\em Phys. Rept.} {\bf
  23} (1976)  245--249}.

\bibitem{'tHooft:1976ac}
G.~'t~Hooft, {\em Proceedings of the EPS International} (1976)  1225.

\bibitem{'tHooft:1981ht}
G.~'t~Hooft,
\href{http://dx.doi.org/10.1016/0550-3213(81)90442-9}{{\em Nucl. Phys.} {\bf
  B190} (1981)  455}.

\bibitem{DeGrand:1980eq}
T.~A. DeGrand and D.~Toussaint,
\href{http://dx.doi.org/10.1103/PhysRevD.22.2478}{{\em Phys. Rev.} {\bf D22}
  (1980)  2478}.

\bibitem{Kronfeld:1987ri}
A.~S. Kronfeld, M.~L. Laursen, G.~Schierholz, and U.~J. Wiese,
\href{http://dx.doi.org/10.1016/0370-2693(87)90910-5}{{\em Phys. Lett.} {\bf
  B198} (1987)  516}.

\bibitem{Kronfeld:1987vd}
A.~S. Kronfeld, G.~Schierholz, and U.~J. Wiese,
\href{http://dx.doi.org/10.1016/0550-3213(87)90080-0}{{\em Nucl. Phys.} {\bf
  B293} (1987)  461}.

\bibitem{Suzuki:2009xy}
T.~Suzuki, M.~Hasegawa, K.~Ishiguro, Y.~Koma, and T.~Sekido,
  \href{http://dx.doi.org/10.1103/PhysRevD.80.054504}{{\em Phys. Rev.} {\bf
  D80} (2009)  054504},
\href{http://arxiv.org/abs/0907.0583}{{\tt arXiv:0907.0583 [hep-lat]}}.

\bibitem{Ezawa:1982bf}
Z.~F. Ezawa and A.~Iwazaki,
\href{http://dx.doi.org/10.1103/PhysRevD.25.2681}{{\em Phys. Rev.} {\bf D25}
  (1982)  2681}.

\bibitem{Amemiya:1998jz}
K.~Amemiya and H.~Suganuma,
  \href{http://dx.doi.org/10.1103/PhysRevD.60.114509}{{\em Phys. Rev.} {\bf
  D60} (1999)  114509},
\href{http://arxiv.org/abs/hep-lat/9811035}{{\tt arXiv:hep-lat/9811035}}.

\bibitem{Bornyakov:2003ee}
V.~G. Bornyakov, M.~N. Chernodub, F.~V. Gubarev, S.~M. Morozov, and M.~I.
  Polikarpov, \href{http://dx.doi.org/10.1016/S0370-2693(03)00368-X}{{\em Phys.
  Lett.} {\bf B559} (2003)  214--222},
\href{http://arxiv.org/abs/hep-lat/0302002}{{\tt arXiv:hep-lat/0302002}}.

\bibitem{Mendes:2008ux}
T.~Mendes, A.~Cucchieri, A.~Maas, and A.~Mihara,
\href{http://arxiv.org/abs/0809.3741}{{\tt arXiv:0809.3741 [hep-lat]}}.

\bibitem{Dudal:2004rx}
D.~Dudal, J.~Gracey, V.~Lemes, M.~Sarandy, R.~F. Sobreiro, S.~P. Sorella, and
  H.~Verschelde, \href{http://dx.doi.org/10.1103/PhysRevD.70.114038}{{\em Phys.
  Rev.} {\bf D70} (2004)  114038},
\href{http://arxiv.org/abs/hep-th/0406132}{{\tt arXiv:hep-th/0406132}}.

\bibitem{Capri:2005tj}
M.~A.~L. Capri, V.~E.~R. Lemes, R.~F. Sobreiro, S.~P. Sorella, and R.~Thibes,
  \href{http://dx.doi.org/10.1103/PhysRevD.72.085021}{{\em Phys. Rev.} {\bf
  D72} (2005)  085021},
\href{http://arxiv.org/abs/hep-th/0507052}{{\tt arXiv:hep-th/0507052}}.

\bibitem{Capri:2006cz}
M.~A.~L. Capri, V.~E.~R. Lemes, R.~F. Sobreiro, S.~P. Sorella, and R.~Thibes,
  \href{http://dx.doi.org/10.1103/PhysRevD.74.105007}{{\em Phys. Rev.} {\bf
  D74} (2006)  105007},
\href{http://arxiv.org/abs/hep-th/0609212}{{\tt arXiv:hep-th/0609212}}.

\bibitem{Capri:2008ak}
M.~A.~L. Capri, V.~E.~R. Lemes, R.~F. Sobreiro, S.~P. Sorella, and R.~Thibes,
  \href{http://dx.doi.org/10.1103/PhysRevD.77.105023}{{\em Phys. Rev.} {\bf
  D77} (2008)  105023},
\href{http://arxiv.org/abs/0801.0566}{{\tt arXiv:0801.0566 [hep-th]}}.

\bibitem{Shinohara:2003mx}
T.~Shinohara, K.~I. Kondo, and T.~Murakami,
  \href{http://dx.doi.org/10.1016/S0920-5632(03)02701-4}{{\em Nucl. Phys. Proc.
  Suppl.} {\bf 129} (2004)  748--750},
\href{http://arxiv.org/abs/hep-lat/0309164}{{\tt arXiv:hep-lat/0309164}}.

\bibitem{Maas:2005xh}
A.~Maas, \href{http://dx.doi.org/10.1016/j.cpc.2006.02.005}{{\em Comput. Phys.
  Commun.} {\bf 175} (2006)  167--179},
\href{http://arxiv.org/abs/hep-ph/0504110}{{\tt arXiv:hep-ph/0504110}}.

\bibitem{Gracey:2005vu}
J.~A. Gracey, {\em JHEP} {\bf 04} (2005)  012,
\href{http://arxiv.org/abs/hep-th/0504051}{{\tt arXiv:hep-th/0504051}}.

\bibitem{Sternbeck:2007ug}
A.~Sternbeck, L.~von Smekal, D.~B. Leinweber, and A.~G. Williams, {\em PoS}
  {\bf LAT2007} (2007)  340,
\href{http://arxiv.org/abs/0710.1982}{{\tt arXiv:0710.1982 [hep-lat]}}.

\bibitem{Cucchieri:2007zm}
A.~Cucchieri, T.~Mendes, O.~Oliveira, and P.~J. Silva,
  \href{http://dx.doi.org/10.1103/PhysRevD.76.114507}{{\em Phys. Rev.} {\bf
  D76} (2007)  114507},
\href{http://arxiv.org/abs/0705.3367}{{\tt arXiv:0705.3367 [hep-lat]}}.

\bibitem{Maas:2007af}
A.~Maas and S.~Olejnik,
  \href{http://dx.doi.org/10.1088/1126-6708/2008/02/070}{{\em JHEP} {\bf 02}
  (2008)  070},
\href{http://arxiv.org/abs/0711.1451}{{\tt arXiv:0711.1451 [hep-lat]}}.

\bibitem{Davydychev:1991va}
A.~I. Davydychev,
\href{http://dx.doi.org/10.1016/0370-2693(91)91715-8}{{\em Phys. Lett.} {\bf
  B263} (1991)  107--111}.

\bibitem{Anastasiou:1999ui}
C.~Anastasiou, E.~W.~N. Glover, and C.~Oleari, {\em Nucl. Phys.} {\bf B572}
  (2000)  307--360,
\href{http://arxiv.org/abs/hep-ph/9907494}{{\tt hep-ph/9907494}}.

\bibitem{Huber:2008mq}
M.~Q. Huber, R.~Alkofer, and K.~Schwenzer, {\em PoS} {\bf Confinement8} (2008)
  174,
\href{http://arxiv.org/abs/0812.4451}{{\tt arXiv:0812.4451 [hep-ph]}}.

\bibitem{Dunne:1987am}
G.~V. Dunne and I.~G. Halliday,
{\em Phys. Lett.} {\bf B193} (1987)  247.

\bibitem{Halliday:1987an}
I.~G. Halliday and R.~M. Ricotta,
{\em Phys. Lett.} {\bf B193} (1987)  241.

\bibitem{Dunne:1987qb}
G.~V. Dunne and I.~G. Halliday,
{\em Nucl. Phys.} {\bf B308} (1988)  589.

\bibitem{Ricotta:1990nd}
R.~Ricotta, in {\em J.J. Giambiagi Festschrift}, p.~350.
\newblock ed. H. Falomir, 1990.

\bibitem{Schwenzer:2008vt}
K.~Schwenzer,
\href{http://arxiv.org/abs/0811.3608}{{\tt arXiv:0811.3608 [hep-ph]}}.

\bibitem{Zwanziger:1992qr}
D.~Zwanziger,
{\em Nucl. Phys.} {\bf B399} (1993)  477--513.

\bibitem{Capri:2008vk}
M.~A.~L. Capri, A.~J. Gomez, V.~E.~R. Lemes, R.~F. Sobreiro, and S.~P. Sorella,
  \href{http://dx.doi.org/10.1103/PhysRevD.79.025019}{{\em Phys. Rev.} {\bf
  D79} (2009)  025019},
\href{http://arxiv.org/abs/0811.2760}{{\tt arXiv:0811.2760 [hep-th]}}.

\bibitem{Min:1985bx}
H.~Min, T.~Lee, and P.~Y. Pac,
\href{http://dx.doi.org/10.1103/PhysRevD.32.440}{{\em Phys. Rev.} {\bf D32}
  (1985)  440}.

\bibitem{Fazio:2001rm}
A.~R. Fazio, V.~E.~R. Lemes, M.~S. Sarandy, and S.~P. Sorella,
  \href{http://dx.doi.org/10.1103/PhysRevD.64.085003}{{\em Phys. Rev.} {\bf
  D64} (2001)  085003},
\href{http://arxiv.org/abs/hep-th/0105060}{{\tt arXiv:hep-th/0105060}}.

\bibitem{Alkofer:2008nt}
R.~Alkofer, M.~Q. Huber, and K.~Schwenzer,
  \href{http://dx.doi.org/10.1016/j.cpc.2008.12.009}{{\em Comput. Phys.
  Commun.} {\bf 180} (2009)  965--976},
\href{http://arxiv.org/abs/0808.2939}{{\tt arXiv:0808.2939 [hep-th]}}.

\bibitem{Gribov:1977wm}
V.~N. Gribov,
{\em Nucl. Phys.} {\bf B139} (1978)  1.

\bibitem{Zwanziger:1990by}
D.~Zwanziger,
\href{http://dx.doi.org/10.1016/0370-2693(91)90876-R}{{\em Phys. Lett.} {\bf
  B257} (1991)  168--172}.

\bibitem{Zwanziger:1991gz}
D.~Zwanziger,
{\em Nucl. Phys.} {\bf B364} (1991)  127--161.

\bibitem{Zwanziger:1993dh}
D.~Zwanziger,
\href{http://dx.doi.org/10.1016/0550-3213(94)90396-4}{{\em Nucl. Phys.} {\bf
  B412} (1994)  657--730}.

\bibitem{vanBaal:1991zw}
P.~van Baal,
{\em Nucl. Phys.} {\bf B369} (1992)  259--275.

\bibitem{Greensite:2004ke}
J.~Greensite, S.~Olejnik, and D.~Zwanziger,
  \href{http://dx.doi.org/10.1103/PhysRevD.69.074506}{{\em Phys. Rev.} {\bf
  D69} (2004)  074506},
\href{http://arxiv.org/abs/hep-lat/0401003}{{\tt arXiv:hep-lat/0401003}}.

\bibitem{Schwenzer:2008fy}
K.~Schwenzer, {\em PoS} {\bf Confinement8} (2008)  177,
\href{http://arxiv.org/abs/0812.3061}{{\tt arXiv:0812.3061 [hep-ph]}}.

\bibitem{Fister:2010ah}
L.~Fister, Diploma thesis, Karl-Franzens-University Graz,
2010.

\bibitem{Fister:2010yw}
L.~Fister, R.~Alkofer, and K.~Schwenzer, {\em Phys. Lett.} {\bf B688} (2010)
  237--243,
\href{http://arxiv.org/abs/1003.1668}{{\tt arXiv:1003.1668 [hep-th]}}.

\bibitem{Alkofer:2003jr}
R.~Alkofer, C.~S. Fischer, H.~Reinhardt, and L.~von Smekal,
  \href{http://dx.doi.org/10.1103/PhysRevD.68.045003}{{\em Phys. Rev.} {\bf
  D68} (2003)  045003},
\href{http://arxiv.org/abs/hep-th/0304134}{{\tt arXiv:hep-th/0304134}}.

\bibitem{Baulieu:1981sb}
L.~Baulieu and J.~Thierry-Mieg,
\href{http://dx.doi.org/10.1016/0550-3213(82)90454-0}{{\em Nucl. Phys.} {\bf
  B197} (1982)  477}.

\bibitem{ThierryMieg:1985yv}
J.~Thierry-Mieg,
\href{http://dx.doi.org/10.1016/0550-3213(85)90562-0}{{\em Nucl. Phys.} {\bf
  B261} (1985)  55}.

\bibitem{Zwanziger:1989mf}
D.~Zwanziger,
\href{http://dx.doi.org/10.1016/0550-3213(89)90122-3}{{\em Nucl. Phys.} {\bf
  B323} (1989)  513--544}.

\bibitem{Capri:2010an}
M.~A.~L. Capri, A.~J. Gomez, M.~S. Guimaraes, V.~E.~R. Lemes, and S.~P.
  Sorella, \href{http://dx.doi.org/10.1088/1751-8113/43/24/245402}{{\em J.
  Phys.} {\bf A43} (2010)  245402},
\href{http://arxiv.org/abs/1002.1659}{{\tt arXiv:1002.1659 [hep-th]}}.

\end{thebibliography}\endgroup

\end{document}